\documentstyle[aps,prd,amstex,times,eqsecnum,epsfig,hyperref]{revtex}

\newcommand{\textstyler}{}
\newcommand{\lbrak}{(\!(}
\newcommand{\rbrak}{)\!)}

\draft

\begin{document}

\title{An excess power statistic for detection of burst sources of
       gravitational radiation}
\author{
Warren G. Anderson$^{(1,2)}$, 
Patrick R. Brady$^{(2,3)}$, 
Jolien D. E. Creighton$^{(2,4)}$
and {\'E}anna {\'E}. Flanagan$^{(5)}$}
\address{
${}^{(1)}$ 
Department of Physical Sciences,
         University of Texas at Brownsville,
         80 Fort Brown,
         Brownsville, Texas 78520\\
${}^{(2)}$ 
Department of Physics,
         University of Wisconsin---Milwaukee,
         P.O. Box 413,
         Milwaukee, Wisconsin 53201\\
${}^{(3)}$ 
Institute of Theoretical Physics,
         University of California,
         Santa Barbara, California 93106-9530\\
${}^{(4)}$ 
Theoretical Astrophysics,
         California Institute of Technology,
         Pasadena, California 91125\\
${}^{(5)}$ 
Cornell University,
         Newman Laboratory,
         Ithaca, New York 14853-5001}
\date{18 August 2000}

\wideabs{\maketitle
\begin{abstract}%
\quad We examine the properties of an excess power method 
to detect gravitational waves in
interferometric detector data.  This method is designed to detect
short-duration ($\lesssim$ 0.5 s) burst signals of unknown waveform, such as
those from supernovae or black hole mergers.  If only the bursts' duration and
frequency band are known, the method is an optimal detection strategy in both
Bayesian and frequentist senses.  It consists of summing the data power over
the known time interval and frequency band of the burst.  If the detector noise
is stationary and Gaussian, this sum is distributed as a $\chi^2$ (non-central
$\chi^2$) deviate in the absence (presence) of a signal.  One can use
these distributions to compute frequentist detection thresholds for
the measured power.  We derive the method from Bayesian analyses and
show how to compute Bayesian thresholds.  More generically, when only upper
and/or lower bounds on the bursts duration and frequency band are known, one
must search for excess power in all concordant durations and bands.  Two
search schemes are presented and their computational efficiencies are
compared.  We find that given reasonable constraints on the effective
duration and bandwidth of signals, the excess power search can be
performed on a single workstation. 
Furthermore, the method can be almost as efficient as matched filtering
when a large template bank is required:
for Gaussian noise the excess power method can detect a source to a distance at
least half of the distance detectable by matched filtering if the product of
duration and bandwidth of the signals is $\alt 100$, and to a much greater
fraction of the distance when the size of the matched filter bank is large.
Finally, we derive generalizations of
the method to a network of several interferometers under the assumption of
Gaussian noise.  However, further work is required to determine the efficiency
of the method in the realistic context of a detector network with non-Gaussian
noise.  
\end{abstract}
}
\narrowtext

\section{Introduction and Summary}
\label{s:introduction}

\subsection{Background and Motivation}

The inspiral, merger, and ringdown of binary black hole systems may be the
most important source of gravitational radiation for detection by the
kilometer-scale interferometric gravitational wave detectors such as
LIGO~\cite{Abramovici.A:1992}, and
VIRGO~\cite{Bradaschia.C:1990,Caron.B:1997}.  The importance of these sources
is twofold \cite{Flanagan.E:1998}:
\begin{enumerate}
\item A large amount of gravitational radiation is expected to be emitted by
the merger of two black holes. For intermediate mass ($\sim 10
M_\odot\mbox{--}1000 M_\odot$) black hole binaries this radiation will be in
the frequency band of highest sensitivity for LIGO and VIRGO.%
\footnote{While the relative abundance of such systems is still a very open 
   question, we are encouraged by two recent developments in the astrophysics
   literature: (i) evidence suggesting that black holes in this mass range may
   exist~\cite{Colbert.E:1999,Ptak.A:1999} and (ii) a globular cluster model
   that suggests LIGO I may expect to see about one black hole coalescence
   event during the first two years of operation~\cite{Zwart.S:1999}.}
These sources should therefore be amongst the brightest in the sky, and
visible to much greater distances than other sources. The detection rate for
coalescing binary black holes could therefore be higher than for any other
source.
\item The radiation emitted from the merger of black holes probes the
strong field regime of a purely gravitational system. This radiation should
therefore provide a sensitive test of general relativity.
\end{enumerate}
These benefits can only be realized, however, if the gravitational radiation
from black hole mergers can be detected.

The best understood and most widely developed technique for detection of
gravitational waves with interferometric detectors is matched
filtering~\cite{300yrs,GRASP}.  Matched filtering is the optimal
technique if the entire waveform to be  
detected is accurately known in advance (up to
a few unknown parameters).  Unfortunately, the gravitational 
radiation from black hole mergers results from highly non-linear
self-interaction of the gravitational field.  This makes it extremely
difficult to obtain gravitational waveforms.  Efforts to do so 
have met with only limited success thus far.  Binary black hole
mergers will therefore not amenable to detection by matched filtering,
at least for the first gravitational wave searches in the 2002-2004
time frame.

Similarly, there are other classes of sources, like core-collapse of
massive stars in supernovae, or the accretion induced collapse of
white dwarfs, for which the physics is too complex to allow
computation of detailed gravitational waveforms.  For these sources,
as for binary black hole mergers, we must seek alternative signal detection
methods.  These methods are often called ``blind search'' methods.

One class of search methods is based on time-frequency decompositions
of detector data.  (For an exploration of a variety of other methods,
see Ref.\ \cite{Arnaud.N:1999}.)
Time-frequency strategies have become standard in
many other areas of signal analysis~\cite{Boashash.B:1992}.  There is
also a growing literature on the application of time-frequency 
methods to gravitational waves~%
\cite{Feo.M:1996A,Feo.M:1996B,Feo.M:1997,Krolak.A:1996,Innocent.J:1997,%
      Goncalves.P:1998,Chassande-Mottin.E:1998,Innocent.J:1998,Arnaud.N:1999,%
      Anderson.W:1999,Mohanty.S:1999}.
For binary black hole mergers, it is possible to
make crude estimates of signals' durations and frequency bands
\cite{Flanagan.E:1998}, although these estimates need to be firmed up
and refined by numerical relativity simulations.  This suggests that
one should look only in the relevant time-frequency window of the
detector output.

Flanagan and Hughes \cite{Flanagan.E:1998} (FH) have suggested a
particular time-frequency method for blind searches.  The
method uses only knowledge of the duration  
and frequency band of the signal: one simply computes the total power
within this time-frequency window, and repeats for different start
times.  The method detects a signal if there is more power 
than one expects from detector noise alone.  Thus, we call it the \emph{excess
power} search method.  Similar methods have been discussed
elsewhere in the gravitational wave literature.  Schutz~\cite{Schutz.B:1991}
investigated the method in the context of the cross-correlation of outputs
from different detectors.  An autocorrelation filter for unrestricted
frequencies was published by Arnaud \textit{et
al.}~\cite{Arnaud.N:1999,Arnaud.N:1999b} shortly after and
independently of FH\@.  A generalization of the 
excess power filter has also been discussed in the signal analysis
literature~\cite{Fawcett.J:1991}, where it has recently been ``attracting
considerable interest''~\cite{Streit.R:1999}.  Finally a method
closely related to the excess power method has recently been explored
by Sylvestre \cite{Sylvestre}.

The excess power method distinguishes itself for the detection of
signals of known duration and frequency band by a single compelling
feature: in the absence of any other knowledge about the signal,
\emph{the method can be shown to be optimal}.  Furthermore, it can 
be shown that for mergers of a sufficiently short duration and narrow
frequency band, it performs nearly as well as matched filtering. 

The essence of the power filter is that one compares the power of the data in
the estimated frequency band and for the estimated duration to the known
statistical distribution of noise power.  It is straightforward to show that if
the detector output consists solely of stationary Gaussian noise, the power in
the band will follow a $\chi^2$ distribution with the number of degrees of
freedom being twice the estimated time-frequency volume (i.e., the product of
the time duration and the frequency band of the signal).  If a gravitational
wave of sufficiently large amplitude is also present in the detector
output, an excess of power will be observed; in this case, the power
is distributed as a non-central $\chi^2$ distribution \cite{Abramowitz.M:1972}
with non-centrality parameter given by the 
signal power.  The signal is detectable if the excess power is much
greater than the fluctuations in the noise power which scales as the 
square-root of the time-frequency volume. Thus, the viability of the excess
power method depends on the expected duration and bandwidth of the
gravitational wave as well as on its intrinsic strength.
For instance, the method is not competitive with matched filtering in
detecting binary neutron star inspirals, since the time-frequency
volume for such signals is very large, $\agt 10^4$.

To implement this method, one needs to decide the range of frequency
bands and durations to search over.  For initial LIGO, the most
sensitive frequency band is $\sim100\mbox{--}300\,\text{Hz}$, and it
makes sense to search just in this band.  For binary black hole
mergers, signal durations might be of order tens or hundreds of milliseconds,
depending on the black hole masses and spins \cite{Flanagan.E:1998}.  
Thus, the time-frequency volume of a merger signal can be as large as
$\sim 100$, and its power would need to be more than one tenth as large as
the noise power for detectability with the excess power method.

One can also establish operational lower bounds on the time durations and
frequency bands of interest.  Because the largest operational frequency
bandwidth is $200\,\text{Hz}$ for the initial LIGO interferometers, the
shortest duration of signal that need be considered is $5\,\text{ms}$ (for a
minimum time-frequency volume of unity).  Similarly, for a maximum duration of
$0.5\,\text{s}$, the smallest bandwidth that needs to be considered is
$2\,\text{Hz}$.  The excess power in any of the allowed bandwidths and
durations can thus be obtained by judiciously summing up power that is output
from a bank of one hundred $2\,\text{Hz}$ band-pass filters (spanning the
$200\,\text{Hz}$ of peak interferometer sensitivity) for the required
duration. 

Having established the statistic and its operational range of parameters, the
following simple algorithm for implementing the excess power method
emerges naturally:
\begin{enumerate}
\item Pick a start time $t_s$, a time duration $\delta t$ (containing $N$
data samples), and a frequency band $[f_s,f_s+\delta f$].
\item Fast Fourier transform (FFT) the block of (time domain) detector data 
for the chosen duration and start time.
\item Sum the power in each of the $\sim$ one hundred 2 Hz bands spanning the 
peak sensitivity region of the detector.
\item Further sum the power in the 2 Hz bands which correspond to the chosen
frequency band.
\item Calculate the probability of having obtained the summed power from
Gaussian noise alone using a $\chi^2$ distribution with $2\times \delta t
\times \delta f$ degrees of freedom.
\item If the probability is significant, record a detection.
\item Repeat the process for all allowable choices of start times
$t_s$, durations $\delta t$, starting frequencies $f_s$ and bandwidths
$\delta f$.
\end{enumerate}
This procedure, which must be repeated for every possible start time, can lead
to moderately-large computational requirements. We find that the computational
efficiency of  this implementation, which we call the short FFT algorithm, can
be improved upon by considering data segments much longer than the longest
signal time duration. In this case, after summing over the chosen band, we
must FFT the data back into the time domain. This implementation, which we
call the long FFT algorithm, is more efficient by at least a factor of $\sim4$
over the parameter space of interest.

The most significant drawback of the filter outlined above is that the
$\chi^2$ statistic is appropriate only to Gaussian noise.  Real
detector noise will contain significant non-Gaussian components.
There are likely to be transient bursts of broad band noise
that have characteristics very similar to black hole merger signals.  

The non-Gaussianity of real detector noise leads us to two considerations.
First, like most blind search methods, the excess power method will
likely be a useful tool for characterizing and investigating the
non-Gaussian components of the noise.
In particular, it can provide a simple and automated procedure for garnering
statistical information about noise bursts.  This is a useful and
important feature of the excess power method, 
even though we focus almost exclusively on signal detection in this
paper. Second, since the method cannot distinguish between noise
bursts and signals in any one detector, it will be essential to use
multiple-detector versions of the power statistic for actual signal 
detections.  In Sec.~\ref{s:multipleDetectors} we derive 
the optimal multi-detector generalization of the excess power statistic
under the assumption of Gaussian noise.  It will be important in the
future to generalize this analysis to allow for (uncorrelated)
non-Gaussian noise components in individual detectors.

The layout of this paper is as follows: in Sec.~\ref{s:overview} we begin with
an overview of the filter and some of its properties. This is done with an eye
toward implementation, so that readers whose primary interest is in
applying the filter need not concern themselves with mathematical aspects of
the statistical theory of receivers.  Subsequently we discuss properties
of the excess power statistic in Sec.~\ref{s:detection}, its
derivation from a Bayesian framework in Sec.~\ref{s:bayes_analysis}, an
efficient implementation of the statistic in Sec.~\ref{s:computation},
and the generalization of the power statistic to multiple detectors in
Sec.~\ref{s:multipleDetectors}.

\subsection{Overview}
\label{s:overview}

The output $h(t)$ of the gravitational wave detector is sampled at a finite
rate $1/\Delta t$ to produce a time series $h_j=h(j\Delta t)$, where
$j = 0,1,2\ldots$.  This output can be written as
\begin{equation}
\label{e:output}
  h_j = n_j + s_j
\end{equation}
where $n_j$ is the detector noise and $s_j$ is a (possibly absent) signal.
For most of this paper we assume that the noise is stationary and Gaussian.
Under these assumptions, the components of the Fourier transform of a segment
containing $N$ samples of noise,
\begin{equation}
\label{e:FourierTransform}
  \tilde{n}_k = {\textstyler{\sum_{j=0}^{N-1}}} n_j e^{2\pi ijk/N},
\end{equation}
can be taken to be independent (we discuss the extent to which this is true in
Sec.~\ref{s:computation}). The one-sided power spectrum $S_k$ of the noise is
defined by: 
\begin{equation}
\label{e:powerSpectrum}
  \langle\tilde{n}_k\tilde{n}_k^\ast\rangle = \case{1}{2} S_k.
\end{equation}
Here, $\langle\cdot\rangle$ indicates an average over the noise distribution
and $\ast$ denotes complex conjugation.

Consider a situation where all possible signals have a fixed time duration
$\delta t$ and are band-limited to a frequency band $[f_s,f_s+\delta
f]$, but no other information is known about them.  Then, as we show in
Sec.~\ref{s:bayes_analysis} 
the \emph{optimal} statistic for detection of this class of signals is the
\emph{excess power statistic}
\begin{equation}
\label{e:excessPower}
  {\mathcal{E}} = 4{\textstyler{\sum_{k_1 \le k < k_2}}} |\tilde{h}_k|^2/S_k.
\end{equation}
The sum in Eq.~(\ref{e:excessPower}) is over the positive frequency components
$k_1 \le k < k_2$ that define the desired frequency band. The number of
frequency components being summed is therefore equal to the time-frequency
volume $V=\delta t\,\delta f=k_2-k_1$.

\begin{figure}
\begin{centering}
\epsfig{file=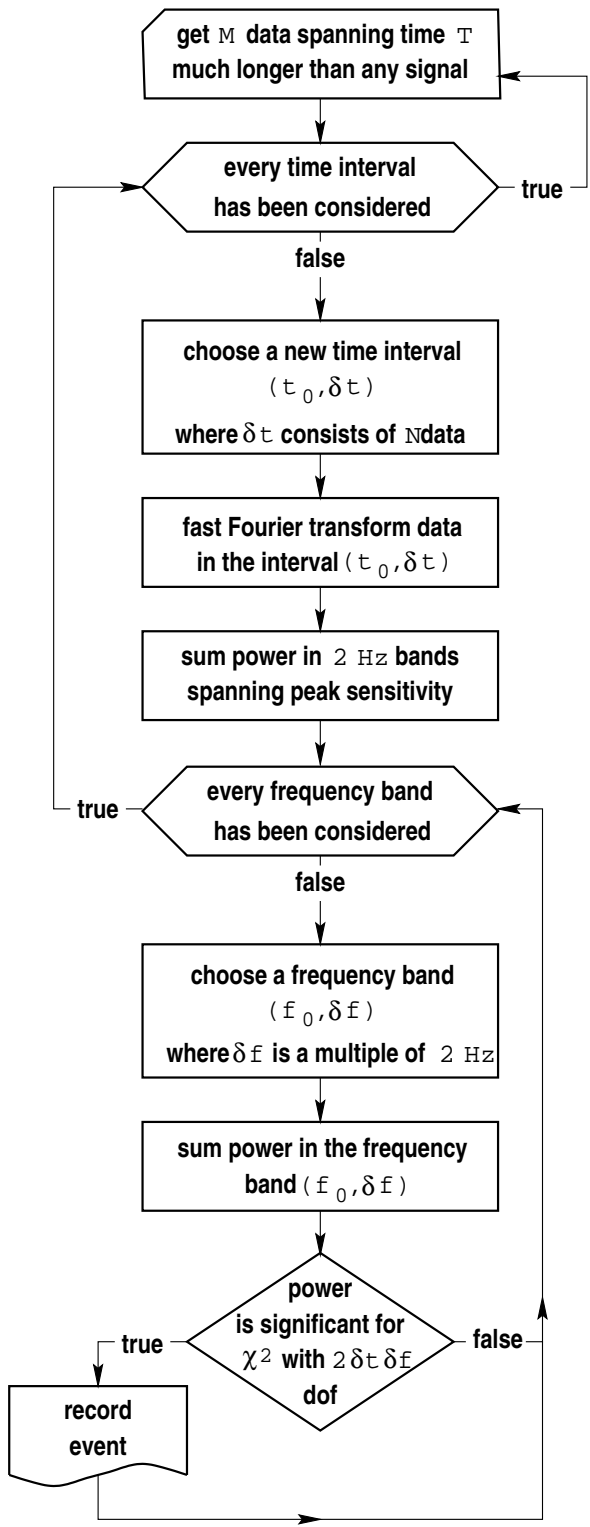,width=0.9\linewidth}
\end{centering}
\vspace{0.3cm}
\caption{\label{f:short_flow}
   A flow chart for the short FFT algorithm for the power filter.}
\end{figure}

\begin{figure}
\begin{centering}
\epsfig{file=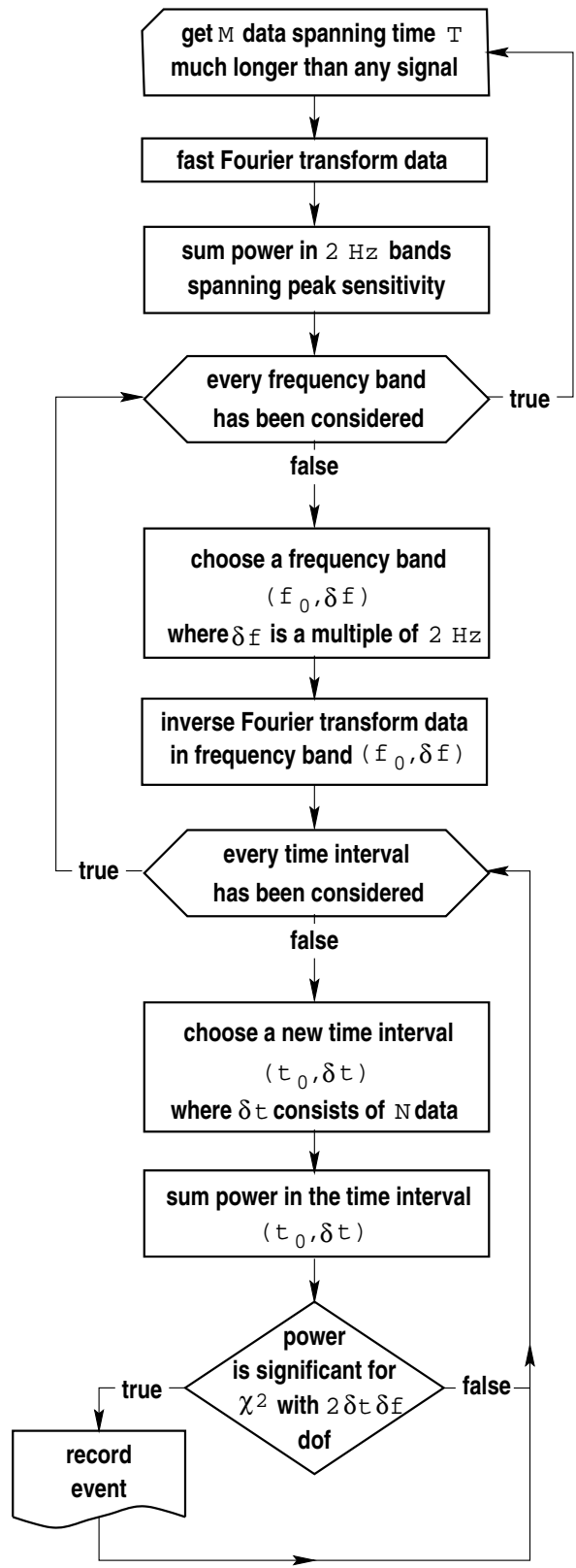,width=0.9\linewidth}
\end{centering}
\vspace{0.3cm}
\caption{\label{f:long_flow}
   A flow chart for the long FFT algorithm for the power filter.}
\end{figure}

In practice, we search over every time-frequency window that is consistent
with a \emph{range} of possible time durations and bandwidths. The number of
such windows per start time is
\begin{equation}
  \label{e:numWindows} N_{\text{windows}} =
  \case{1}{2}N_{\text{channels}}(N_{\text{channels}}+1)
  (N_{\text{max}}-N_{\text{min}}+1)
\end{equation}
where $N_{\text{max}}=\delta t_{\text{max}}/\Delta t$ is the number of samples
in the longest expected signal duration $\delta t_{\text{max}}$, and
$N_{\text{min}}=\delta t_{\text{min}}/\Delta t$.  
Here $N_{\text{channels}} = \delta f_{\text{max}} / \delta f_{\text{min}}$ is
the ratio of the largest bandwidth searched over $\delta f_{\text{max}}$
to the shortest bandwidth $\delta f_{\text{min}}$.
One strategy for this search was outlined above.  A flowchart for this
algorithm is presented in Fig.~\ref{f:short_flow}.  We call this algorithm the
\emph{short FFT method}.  We have also considered a second algorithm which we
call the \emph{long FFT method}, and its flowchart is shown in
Fig.~\ref{f:long_flow}. 

Under the conditions that (i) the number $M$ of time domain data points being
filtered is large and (ii) one searches over many different time-frequency
volumes, the long FFT method is computationally more efficient than the short
FFT method: the long FFT eliminates the redundancy of Fourier transforming
data more than once when it falls into overlapping time-frequency volumes. In
Sec.~\ref{s:computation} we estimate the computational costs of the two
methods. We find that the short FFT method requires
\begin{equation}
\label{e:costShort}
  C_{\text{short}}\simeq \frac{V^2}{2\alpha_{\text{max}}} \left(
  \frac{3\log_2 V}{\alpha_{\text{max}}} + \frac{V}{3} \right)
\end{equation}
floating-point operations per start time where
$V=\delta t_{\text{max}}\,\delta f_{\text{max}}$ is the maximum time-frequency
volume and $\alpha_{\text{max}}=\delta f_{\text{max}}\Delta t$ is the maximum
dimensionless bandwidth. (The frequency band from DC to Nyquist corresponds to
$\alpha_{\text{max}}=\case{1}{2}$.)  The long FFT method requires only
\begin{equation}
\label{e:costLong}
  C_{\text{long}}\simeq \alpha_{\text{max}}^{-1}V^2\ln V
\end{equation}
floating-point operations per start time; the computational improvement is a
factor of
\begin{equation}
\label{e:costComparison}
  \frac{C_{\text{short}}}{C_{\text{long}}} \sim
  \frac{3}{2\ln2}\frac{1}{\alpha_{\text{max}}} + \frac{1}{6}\frac{V}{\ln V}.
\end{equation}
The first term shows that there is at least a factor of $\sim4$ to be gained
by the long FFT method; in addition to this, the computational gain increases
with the total time-frequency volume to be searched.  For $V=100$, the value
of the second term is also $\sim4$.

Having computed the total power for the various
time-frequency windows of interest, one must decide which, if any, of
those windows might contain a signal. 
A signal increases the expected
power in a window, so we seek windows containing a statistically
significant excess of power.  If one knows the distribution of the
noise data, one can derive the distribution for the noise power and thus
set detection thresholds on the power.

The power $\mathcal{E}$ is distributed as a $\chi^2$ distribution with $2V$
degrees of freedom in the absence of a signal.  When a signal is present,
$\mathcal{E}$ is distributed as a non-central $\chi^2$ distribution with $2V$
degrees of freedom \cite{Abramowitz.M:1972}, where the non-central
parameter is the signal power $A^2$:
\begin{equation}
\label{e:signalPower}
  A^2 = 4{\textstyler{\sum_{k_1\le k < k_2}}} |\tilde{s}_k|^2/S_k.
\end{equation}
The quantity $A$ also represents the signal-to-noise ratio
that one would expect to achieve if a matched filter were used to
detect the signal.   

\begin{figure}
\epsfig{file=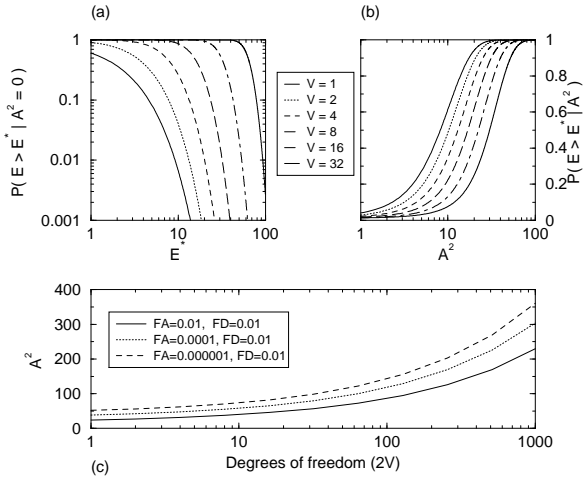,width=0.9\linewidth}
\vspace{0.3cm}
\caption{\label{f:chi2}
  Cumulative probability functions for (a) the $\chi^2$ distribution and
  (b) the non-central $\chi^2$ distribution for various degrees of freedom
  $2V$.  The curves in (a) give the probability that the power
  $\mathcal{E}$ exceeds a threshold ${\mathcal{E}}^\star$ when no signal is
  present.  They can be used to fix the threshold given the time-frequency
  volume $V$ and the desired false alarm probability. The curves in (b)
  give the probability of detecting a signal whose power is $A^2$ in the
  time-frequency volume $V$ and given a threshold
  ${\mathcal{E}}^\star$, where the threshold ${\cal E}^\star$ is chosen to give
  a false alarm probability of 0.01.  (c) The
  signal-to-noise ratio $A^2$ necessary to achieve a given false alarm
  probability (FA) and a false dismissal probability (FD) of 0.01 for
  various values of the number $2V$ of degrees of freedom.}
\end{figure}

In Fig. \ref{f:chi2} we show the central and non-central $\chi^2$
distributions for several choices of parameters.  It is
straightforward to use the $\chi^2$ distributions plotted in 
Fig.~\ref{f:chi2}~(a) to set a frequentist threshold for the excess power
statistic so that a desired false alarm probability is achieved; then the
false dismissal probability can be computed as a function of signal amplitude
using the non-central $\chi^2$ distributions plotted in Fig.~\ref{f:chi2}~(b).
Alternatively, on can fix both false alarm and false dismissal probabilities,
and then use the $\chi^2$ and non-central $\chi^2$ distributions to determine
the expected signal-to-noise ratio ($A$) of the signal which achieves these
probabilities.  A curve demonstrating this for a false alarm probability of
$10^{-9}$ and and false dismissal probability of $0.01$ is show in
Fig.~\ref{f:chi2}~(c).

A derivation of the statistical properties of the excess power
statistic (both with and without a signal) can be found in
Sec.~\ref{s:detection}.  In Sec.\ \ref{s:bayes_analysis} we show that
the method is optimal and unique under suitable assumptions using a
Bayesian analysis.

\begin{figure}
\begin{centering}
\epsfig{file=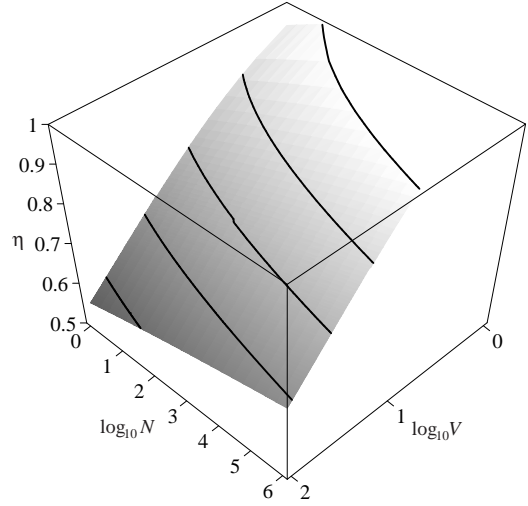,width=0.9\linewidth}
\end{centering}
\caption{\label{f:eta}
  The relative effectiveness $\eta$ of the excess power method with
  respect to matched filtering, for the case where the time-frequency
  window is known in advance, as a function of
  the effective number of independent templates ${\cal 
  N}_{\text{eff}}$ characterizing the space of signals being sought,
  and the time-frequency volume $V$.  A false alarm probability
  corresponding to one false alarm per one hundred days of observation
  (taken to have $1.728\times10^9$ independent
  arrival times), and a correct detection probability of $0.99$
  were assumed.  The quantity $\eta$ is 
  the ratio of the minimum signal amplitude that is required to achieve these
  false alarm and dismissal probabilities for the excess power method, to
  the corresponding minimum amplitude for the matched filtering method.
  The loss in event rate due employing the
  excess power method rather than matched filtering is $\eta^3$.  This
  plot can be generated by combining Eqs.\ (\protect{\ref{e:Amindef}}),
  (\protect{\ref{e:AEPdef}}), (\protect{\ref{e:AMFdef}}) and
  (\protect{\ref{e:eta}}) of   the text.  }
\end{figure}

The excess power statistic requires minimal information about the
signals to be detected, making it a useful statistic for poorly modeled
sources.
Nevertheless, it is useful to compare the detection
efficiency of the method to that of matched filtering.  
We characterize the excess power filter by the time-frequency volume
$V$ and the matched filter bank by the number $\mathcal{N}_{\text{eff}}$ of
effectively independent filters.  For given false alarm and false
dismissal probabilities, we denote the 
amplitude of the weakest signals detectable using the excess power filter or a
bank of matched filters by $A_{\text{min}}^{\text{EP}}$ and
$A_{\text{min}}^{\text{MF}}$ respectively.  The \textit{relative
effectiveness} $\eta$ of the two search methods is given by the ratio of these
amplitudes: $\eta=A_{\text{min}}^{\text{EP}}/A_{\text{min}}^{\text{MF}}$.  In
other words, the excess power statistic can detect a source at a fraction
$\eta$ of the distance to which a bank of matched filters can.  The relative
effectiveness $\eta$ is plotted as a function of the time-frequency volume $V$
and the number of templates $\mathcal{N}_{\text{eff}}$ is shown in
Fig.~\ref{f:eta}.  This figure shows that for time frequency
volumes $V$ less than $\sim 100$, and for all values of the effective number of
templates ${\cal N}_{\text{eff}}$, the relative effectiveness $\eta$ is
greater than $1/2$.

\section{The search method in a Single Interferometer}
\label{s:detection}

In this section we define the search method in the context of a single
interferometer, and derive its operating characteristics from 
the frequentist statistical framework.

\subsection{Definition of method for a single time-frequency window}
\label{s:definition}

Consider stretches of discretely sampled detector data 
${\bf h} = \{h_0, h_1, \ldots,h_{N-1} \}$ consisting of
$N$ data points.  We will denote by $\mathcal{V}$ the
$N$-dimensional vector space of all such data stretches.
We assume that the detector output consists of a stationary, zero-mean,
Gaussian noise component $n_j$, plus a possible signal $s_j$, so that
$h_j=n_j+s_j$.  Under these assumptions, the statistical properties of
the noise are characterized by the $N \times N$ correlation matrix
\begin{equation}
R_{ij} \equiv  \langle n_in_j\rangle = C_n(|i-j|\Delta t).
\label{e:correlationFunction}
\end{equation}
Here $C_n(t)$ is the correlation function of the noise and $\Delta t$
is the sampling time.  This correlation matrix
determines a natural inner product on $\mathcal{V}$ given by 
\begin{equation}
\label{e:innerProduct}
({\mathbf{a}},{\mathbf{b}}) = {\textstyler{\sum_{i,j=0}^{N-1}}} a_i Q_{ij}
b_j .
\end{equation}
where ${\bf Q} = {\bf R}^{-1}$.

We now discuss the notion of time-frequency projections.
Consider the time-frequency window 
\begin{equation}
{\cal T} = \{t_s, \delta t, f_s, \delta f \}
\end{equation}
defined by the 
frequency interval $[f_s, f_s + \delta 
f]$, where $f_s$ is a starting frequency and $\delta f$ is a 
bandwidth,
and the 
time interval $[t_s , t_s + \delta t]$, where $t_s$ is a starting
time and $\delta t$ is a duration.  
Suppose that we want to focus attention on that portion of the data
that lies inside the time-frequency window ${\cal T}$, to the extent
that this is meaningful.  One obvious thing to do is to truncate the
data in the time domain, perform a discrete Fourier transform into the
frequency domain, and then throw away the data points outside the
frequency band of interest. 
One then obtains the quantities 
\begin{equation}
{\tilde H}_K = \sum_{J=0}^{N_t-1} e^{2\pi iJK/N_t} h_{j+J},
\label{e:methodI2}
\end{equation}
where $j = t_s/\Delta t$, $N_t = \delta t / \Delta t$, and $K$ runs
over the range $f_s \delta t \le K \le (f_s + \delta f) \delta t$.  
We denote by ${\cal W}_{\cal T}$ the vector space of the projected
data ${\tilde H}_K$.  The dimension of this vector space over the
real numbers is
\begin{equation}
{\mathrm{dim}} \, {\cal W}_{\cal T} = 2 \delta t \delta f = 2 V_{\cal T},
\label{e:TFvol}
\end{equation}
where $V_{\cal T} \equiv \delta t \delta f$ is the time-frequency
volume of the time-frequency window ${\cal T}$.

Of course, there are many other methods of attempting to pick out the
portion of the data in the time-frequency window ${\cal T}$.%
\footnote{For example, one could FFT the entire data segment, truncate
it in the frequency domain, FFT back to the time domain, and then
truncate it again in the time domain.}
The lack of a preferred unique method is due to the uncertainty principle. 
In many circumstances the differences between different reasonable choices
will be relatively unimportant.  For the remainder of this section we
will assume that we have picked some reasonable projection
method.%
\footnote{The choice of a 
projection method corresponds mathematically to the choice of a $2
V_{\cal T}$-dimensional subspace of the dual space ${\cal V}^*$ of
${\cal V}$.  When one specifies in addition the detector noise
spectrum, the projection method determines a $2 V_{\cal T}$
dimensional subspace of ${\cal V}$.}
We can write the
projected data in general as
\begin{equation}
{\bar h}_J = \sum_{j=0}^{N-1} A_{J}^{\ j} \ h_j,
\label{e:mapp}
\end{equation}
where $A_J^j$ is a real $2V_{\cal T} \times N$ matrix, 
the quantities ${\bar h}_J$ are real, and $J$ runs over $0 \le J \le 2
V_{\cal T} -1$.

We define the power statistic associated with ${\cal T}$ and with a
choice of projection method to be
\begin{equation}
{\cal E}_{\cal T}({\bf h}) \equiv \sum_{I,J=0}^{2 V_{\cal T}-1} \,
Q_{IJ} {\bar h}_I {\bar h}_J,
\label{e:calEdef}
\end{equation}
where $\sum_J Q_{IJ}R_{JK}= \delta_{IK}$ and 
\begin{equation}
R_{JK} = \langle {\bar h}_J {\bar h}_K \rangle = \sum_{j,k} \ A_J^{\
j} A_K^{\ k} R_{jk}
\label{e:correl1}
\end{equation}
is the correlation matrix of the projected data.  
The quantity $\mathcal{E}$ is, roughly speaking, just the total power in the
data stream within the given time-frequency window, where power is not the
physical power but is measured relative to the detector noise (i.e.,
it is the conventional power of the pre-whitened data stream).

The statistic can also be described geometrically as follows.  The
linear mapping defined by Eq.\ (\ref{e:mapp}) has a kernel $\left\{ h_i
\right| {\bar h}_I =0 {\mbox{ for all }} I \}$.  The set of all
vectors ${\bf h}$ perpendicular to all elements of this kernel 
with respect to the inner product (\ref{e:innerProduct}) 
form 
a subspace ${\cal V}_{\cal T}$ of ${\cal V}$ 
which can be naturally identified with ${\cal W}_{\cal T}$.  Any
element ${\bf h}$ in ${\cal V}$ can be decomposed as
\begin{equation}
{\bf h} = {\bf h}_\parallel + {\bf h}_\perp,
\label{e:decompose}
\end{equation}
where ${\bf h}_\parallel$ lies in ${\cal V}_{\cal T}$ and ${\bf
h}_\perp$ is perpendicular to all elements of ${\cal V}_{\cal T}$.
The statistic (\ref{e:calEdef}) is the squared norm of the parallel
component: 
\begin{equation}
{\cal E}_{\cal T}({\bf h}) = \left( {\bf h}_\parallel, {\bf
h}_\parallel \right).
\label{e:calEdef1}
\end{equation}

For the simple time-frequency truncation method (\ref{e:methodI2})
discussed above, one can obtain a simple approximate formula for the
statistic.  The correlation matrix of the quantities ${\tilde H}_J$ of Eq.\
(\ref{e:methodI2}) is given by, to a first, crude approximation,
\begin{eqnarray}
\label{e:methodIRval0}
\langle {\tilde H}_J \, {\tilde H}_K \rangle &=&0, \\
\mbox{} \langle {\tilde H}_J \, {\tilde H}_K^* \rangle &=& \case{1}{2}
\delta_{JK} \, S_K,
\label{e:methodIRval}
\end{eqnarray}
where $0 \le J,K \le N_t/2$, 
\begin{equation}
S_K = \delta t \, S_h(K / \delta t) / (\Delta t)^2
\label{e:spectrumdef}
\end{equation}
 and $S_h(f)$ 
is the conventional one-sided power spectral density of the detector
noise.  The expressions (\ref{e:methodIRval0}) and (\ref{e:methodIRval})
are accurate only when $f_s \delta t \gg 1$ and 
when $S_h(f)$ does not vary substantially on scales $\sim 1/\delta t$;
more accurate expressions can be computed if desired from Eqs.\
(\ref{e:correlationFunction}) and (\ref{e:methodI2}).
The definition (\ref{e:calEdef}) now yields
\begin{equation}
{\cal E}_{\cal T}({\bf h}) \approx 4 \sum_{K = f_s
\delta t}^{(f_s + \delta f) \delta t}  | {\tilde H}_K | ^2 / S_K,
\label{e:calEapprox}
\end{equation}
cf., Eq.\ (\ref{e:excessPower}) of the Introduction.  We show in
Sec.\ \ref{s:computation} below that the expression
(\ref{e:calEapprox}) is an adequate approximation to ${\cal E}_{\cal
T}({\bf h})$ for most purposes.

The search method consists of searching over time-frequency windows
${\cal T}$, and selecting as possible events only those windows ${\cal
T}$ for which ${\cal E}_{\cal T}$ exceeds a suitable threshold ${\cal
E}^\star$.  We discuss further how to search over time-frequency windows in
Sec.\ \ref{s:computation} below.  For the remainder of this section we
assume that the time-frequency window ${\cal T}$ is fixed and known,
and discuss the performance of the statistic.

\subsection{Operating characteristics of the statistic}
\label{ss:operation}

\begin{figure}
\epsfig{file=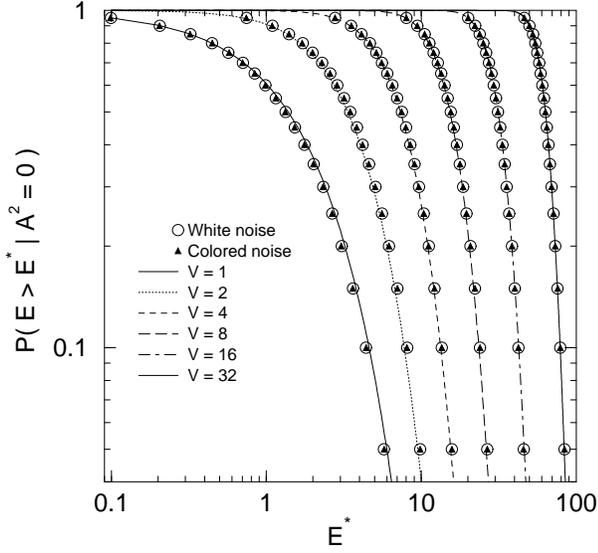,width=0.9\linewidth}
\caption{\label{f:falseAlarm}%
  The probability $P({\mathcal{E}}>{\mathcal{E}}^\star|0)$ of obtaining
  a value of ${\mathcal{E}}$ greater than a given threshold
  ${\mathcal{E}}^\star$, for stationary Gaussian noise in the absence of a
  signal, as a function of ${\mathcal{E}}^\star$.  The quantity
  ${\mathcal{E}}$ has a $\chi^2$ distribution with $2V$ degrees of freedom,
  where $V$ is the time-frequency volume (\protect{\ref{e:TFvol}}) of
  the signal being sought.  The 
  lines show theoretical curves generated according to the standard $\chi^2$
  formula \protect{(\ref{e:falseAlarm})}.  The data points represent values
  obtained from Monte Carlo simulations (described in Sec.\
  \protect{\ref{sss:FourierBasis}}), where the noise is Gaussian and white
  (circles) or colored (triangles).  
  }
\end{figure}

When a signal is not present in the data stream, the statistic
${\cal E} \equiv \mathcal{E}_{\cal T}({\bf h})$
is the sum of the squares of $2V$ independent, zero-mean, unit-variance
Gaussian random variables.%
\footnote{To see this, note that there is a basis of ${\mathcal{W_T}}$ in
  which the correlation matrix $R_{JK}$ is equal to the $2V \times 2V$
  identity matrix, and use Eqs.\ (\protect{\ref{e:calEdef}}) and
  (\protect{\ref{e:correl1}}).}
Thus $\mathcal{E}$ follows a $\chi^2$ distribution with $2V$ degrees of
freedom; the upper-tail cumulative probability is
\begin{equation}
\label{e:falseAlarm}
  Q_0({\mathcal{E}}^\star) = P({\mathcal{E}}>{\mathcal{E}}^\star)
  = \frac{\Gamma(V,{\mathcal{E}}^\star/2)}{\Gamma(V)},
\end{equation}
where $\Gamma(a,x)=\int_x^\infty e^{-t}t^{a-1}dt$ is the incomplete Gamma
function.  The quantity $Q_0({\mathcal{E}}^\star)$ is the \emph{false alarm}
probability for the detection threshold ${\mathcal{E}}^\star$.
The distribution (\ref{e:falseAlarm}) is plotted in Fig.\
\ref{f:falseAlarm} for several values of $V$.
An approximate expression for $Q_0$ in the regime $Q_0 \ll 1$
is \cite{Abramowitz.M:1972}
\begin{eqnarray}
Q_0({\cal E}^\star) &=& \sqrt{2 V \over \pi} {1 \over A_\star^2} \left( 1 +
{A_\star^2 \over 2 V} \right)^V e^{- A_\star^2/2} \nonumber \\
\mbox{} && \times \left[ 1 + O\left({1 \over V} \right) 
+ O\left({1 \over A_\star^2} \right) 
+ O\left({V \over A_\star^4} \right) \right],
\label{e:threshold_approximate}
\end{eqnarray}
where $A_\star^2 \equiv {\cal E}^\star - 2 V$.


We next consider the case when a signal is present, so that ${\bf h} =
{\bf n} + {\bf s}$.  The formula for the statistic ${\cal E}$
given by Eqs.\ (\ref{e:mapp}) and  (\ref{e:calEdef}) becomes
\begin{equation}
{\cal E} = \sum_{I,J=0}^{2V-1} Q_{IJ}
({\bar n}_I + {\bar s}_I) ({\bar n}_J + {\bar s}_J),
\label{e:calE_signal}
\end{equation}  
where ${\bar n}_I = \sum_j A_I^{\ j} n_j$ is the projected noise and
${\bar s}_I = \sum_j A_I^{\ j} s_j$ is the projected signal.
We define the amplitude $A$ of the signal by\footnote{Note that $A$ is
the signal-to-noise ratio that would be obtained by matched filtering
if prior knowledge of the waveform shape allowed one to perform
matched filtering, and if the signal ${\bf s}$ were confined to the
time-frequency window ${\cal T}$.}
\begin{equation}
A^2 = \left( {\bf s}_\parallel, {\bf s}_\parallel \right) =
\sum_{I,J=0}^{2V-1} Q_{IJ} {\bar s}_I {\bar s}_J,
\end{equation}
where we use the notation of Eq.\ (\ref{e:decompose}).  
The expression (\ref{e:calE_signal}) can be simplified by (i) choosing the
basis of ${\cal W}_{\cal T}$ so that $Q_{IJ}=\delta_{IJ}$
(which is roughly equivalent to whitening the detector output for a
large class of time-frequency windows) and (ii) further specializing
the choice of basis so that 
the signal vector is $({\bar s}_0,{\bar s}_1,\ldots,{\bar
s}_{2V-1})=(A,0,\ldots,0)$.  The result is 
\begin{equation}
\label{e:calEsimple}
  {\mathcal{E}} = (\bar{n}_0 + A)^2 + {\textstyler{\sum_{I=1}^{2V-1}}}
  \bar{n}_I^2
\end{equation}
where $\bar{n}_0,\bar{n}_1,\ldots,\bar{n}_{2V-1}$ are independent Gaussian
random variables with zero mean and unit variance.  

From Eq.\ (\ref{e:calEsimple}) one can compute the moment generating
function for the random variable $\mathcal{E}$.  The result is
\begin{equation}
\label{e:momentGeneratingFunction}
  \langle e^{t{\mathcal{E}}}\rangle =
  \frac{\exp[A^2t/(1-2 t)]}{(1-2 t)^V}  .
\end{equation}
The probability distribution for $\mathcal{E}$ can be now obtained by
taking the inverse Laplace transform, which is 
accomplished by expanding the argument of the exponential in Eq.\
(\ref{e:momentGeneratingFunction}) as a power series in $A$.  The result is a
weighted sum of $\chi^2$ probability distribution functions:
\begin{equation}
\label{e:excessPowerProbabilityDistribution}
  p({\mathcal{E}}| A,V) = \sum_{n=0}^\infty
  \frac{e^{-A^2/2}(A^2/2)^n}{n!}
  \frac{e^{-{\mathcal{E}}/2}({\mathcal{E}}/2)^{n+V-1}}{\Gamma(n+V)}.
\end{equation}
This is the non-central $\chi^2$ probability distribution with
non-centrality parameter $A^2$ discussed in Sec.~26.4 of
Ref.~\cite{Abramowitz.M:1972}.  A closed form expression for the
probability distribution is \cite{Groth}
\begin{equation}
\label{e:excessPowerProbabilityDistribution1}
  p({\mathcal{E}}| A,V) = \case{1}{2} e^{-({\cal E} + A^2) /2}
  ( {\cal E}^{1/2}/A )^{V-1} I_{V-1}(A {\cal E}^{1/2}),
\end{equation}
where $I_n(x)$ is the modified Bessel function of the first kind of
order $n$.

\begin{figure}
\epsfig{file=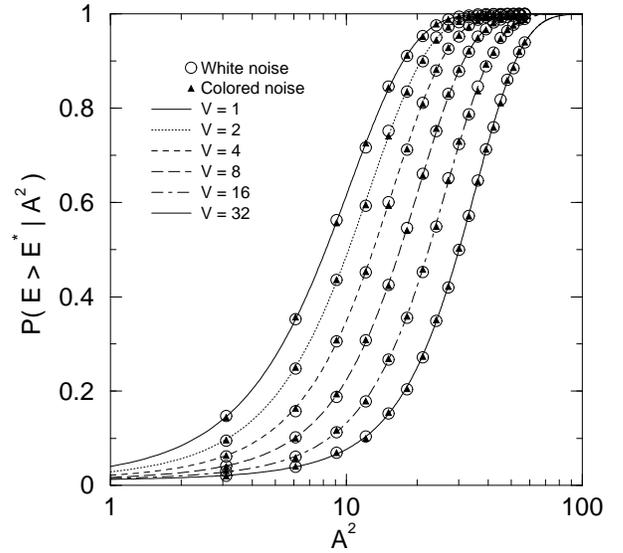,width=0.9\linewidth}
\caption{\label{f:trueDetect}%
  The probability $P({\mathcal{E}}>{\mathcal{E}}^\star|A)$ of obtaining a
  value of ${\mathcal{E}}$ greater than or equal to a given threshold
  ${\mathcal{E}}^\star$ given the presence of a signal of power $A^2$ and
  stationary Gaussian noise, for several different values of the number $2V$
  of degrees of freedom, as a function of $A^2$.  For each $V$, the value of
  ${\mathcal{E}}^\star$ which gives a false alarm probability of $Q_0=0.01$ is
  used.  The lines show theoretical curves generated according to the standard
  non-central $\chi^2$ formula \protect{(\ref{e:trueDetection})}. The data
  points represent values obtained from Monte Carlo simulations
  [described in Sec.\ \protect{\ref{sss:FourierBasis}}], where the
  noise is white (circles) or colored (triangles).}
\end{figure}

The upper-tail cumulative probability distribution for ${\cal E}$
\begin{equation}
\label{e:trueDetection}
  Q_A({\mathcal{E}}^\star,A,V) = P({\mathcal{E}}>{\mathcal{E}}^\star|A,V)
  = \int_{{\mathcal{E}}^\star}^\infty p({\mathcal{E}}| A,V)
  d{\mathcal{E}}
\end{equation}
is the \emph{true detection} probability for a given threshold
${\mathcal{E}}^\star$ and a given signal amplitude $A$.
Figure \ref{f:trueDetect} shows this true detection probability $Q_A$,
expressed as a function of signal strength $A$ and false
alarm probability $Q_0$ [via Eq.\ (\ref{e:falseAlarm})],
evaluated at $Q_0 = 0.01$, for several different values of the
time-frequency volume $V$.

Some qualitative insight into the detectability of a signal can be obtained
from the first two moments of the distribution for $\mathcal{E}$.  The
expected (mean) value is $\langle{\mathcal{E}}\rangle=2V+A^2$, while the
variance is
${\mathrm{Var}}\,{\mathcal{E}}=\langle{\mathcal{E}}^2\rangle
 -\langle{\mathcal{E}}\rangle^2=4V+4A^2$.  For large values of $V$ the
probability distributions are nearly Gaussian within a few sigma of
the expected value, so we can imagine setting the
threshold ${\cal E}^\star$ to be a few times $\sqrt{4V}$ above the
mean noise level $2V$ in order to achieve the required false alarm
probability.  Thus, a signal will be detectable when
${\cal E} - 2 V>(\text{a few})\times \sqrt{4V}$.  In other words, the
signal power $A^2$ can be {\it small} compared to the mean noise power
$2 V$ and still be 
detectable; it need only be comparable to the much smaller
fluctuations $\sim \sqrt{4V}$ in the noise power.


Once one specifies the time-frequency volume $V$ and desired values of
the false alarm probability $Q_0$ 
and true detection probabilities $Q_A$, there is a minimum signal
amplitude $A_{\text{min}}$ that can be detected with the
excess power method.   To compute this amplitude, one first inverts
Eq.~(\ref{e:falseAlarm}) to obtain the required threshold
${\mathcal{E}}^\star$ as a function of $Q_0$ and $V$: ${\cal E}^\star = {\cal
E}^\star(Q_0,V)$.  Second, one inverts Eq.~(\ref{e:trueDetection}) to
obtain the amplitude $A$ as a function of ${\cal E}^\star$,
$Q_A$ and $V$: $A = A({\cal E}^\star,Q_A,V)$.  The minimum signal amplitude is then
given by
\begin{equation}
A_{\text{min}}(Q_0,Q_A,V) = A[{\cal E}^\star(Q_0,V),Q_A,V].
\label{e:Amindef}
\end{equation}
This quantity is plotted as a function of the time frequency volume in
Fig.\ \ref{f:chi2} (c), for various values of $Q_0$ and for $Q_A = 0.99$.


\subsection{Comparison of performance to that of matched filtering.}
\label{s:comparison-MF}

The performance of the excess power statistic should be compared with that of
a matched filtering search for the same class of signals.  
Of course, matched filtering searches will not be possible for
the classes of signals we are interested in (for example supernovae)
due to the lack of theoretical templates; nevertheless the comparison
is useful as benchmark of the excess power method.

We start by discussing how the performance of matched filtering
depends on the set of signals being searched for.
Suppose that a given class of signals 
have a known duration and frequency band, so that they all lie inside
a fixed time-frequency window ${\cal T}$ with time-frequency volume
$V$.  Let $A_\star(Q_0)$ be the signal-to-noise ratio threshold for the
matched filtering search necessary to give a false alarm probability 
of $Q_0$ for each starting time.  Now if the bank of matched filters
consisted of ${\cal N}$ statistically independent filters, then the
threshold $A_\star$ would be given by the formula
\begin{equation}
{\mathrm{erfc}}( A_\star/ \sqrt{2}) = 1 - (1
- Q_0)^{1/{\cal N}} \approx {Q_0/{\cal N}}.
\label{e:th_ideal}
\end{equation}
In the more realistic case where the templates are not all statistically
independent, the formula (\ref{e:th_ideal}) motivates us to define an
effective number of independent templates ${\cal N}_{\text{eff}} = {\cal
N}_{\text{eff}}(Q_0)$ by the relation\footnote{In Ref.\
\protect{\cite{Flanagan.E:1998}} the quantities $A_\star$, $Q_0$ and 
${\cal N}_{\text{eff}}$ were denoted $\rho_\star$,
$\epsilon/{\cal N}_{\text{start-times}}$, and ${\cal N}_{\text{shapes}}$,
respectively.}
\begin{equation}
{\mathrm{erfc}}[{A_\star(Q_0)/\sqrt{2}}] = {Q_0/{\cal N}_{\text{eff}}(Q_0)}.
\label{e:th_ideal2}
\end{equation}
This effective number of templates depends on
the false alarm probability $Q_0$, or, equivalently, on the
detection threshold $A_\star$ via the relation $A_\star = A_\star(Q_0)$.  
For a given class of signals (e.g., inspiralling binaries), it should be
possible to estimate ${\cal N}_{\text{eff}}$ by a modification of the method of
Ref.\ \cite{Owen.B:1996} wherein one does not eliminate the intrinsic
parameters or the signal amplitude and one chooses a minimal match of
order $\sim 0.3$ say to give 
approximately statistically independent templates.  
We suspect that ${\cal N}_{\text{eff}}$ will not differ too
significantly from the actual number of templates used in a
search.%
\footnote{The grid of search templates used will be determined
by having a minimal match \protect{\cite{Owen.B:1996}} of $\sim 0.97$
say instead of $\sim 0.3$, which tends to make the actual number of
templates larger than ${\cal N}_{\text{eff}}$.  On the other hand, a
template grid needs only 2 templates to cover all possible signal
amplitudes and phases (when the other parameters are fixed), whereas
the number of statistically independent templates that can be
generated by varying the amplitude and phase can be much greater than
2 for small $Q_0$.}
In any case, for a given matched filtering search, the detection
threshold and the resulting effective number of independent filters will be
determined by Monte Carlo simulations.

An illustrative special case of matched filtering is when the signal
manifold is a linear subspace ${\cal S}$ of the space of all possible
signals.  In this case the maximum over all templates of the
signal-to-noise ratio squared is simply $\left( {\bf h}_\parallel,
{\bf h}_\parallel \right)$, where ${\bf h}_\parallel$ is the
perpendicular projection of the detector output ${\bf h}$ into ${\cal
S}$ \cite{flanagan.e:1998a}.  Since this quantity has a $\chi^2$
distribution, it follows from the approximate formula
(\ref{e:threshold_approximate}) and from the definition (\ref{e:th_ideal2}) 
that for this special case we have
\begin{eqnarray}
{\cal N}_{\text{eff}}(A_\star) &=&
{\sqrt{2 d} \over A_\star} \left[ 1 + {A_\star^2
\over d} \right]^{d/2} \nonumber \\
\mbox{} &\times& 
\left[ 1 + O\left({1 \over d}\right) +
O\left({1 \over A_\star^2}\right) + O\left( {d \over A_\star^4}\right) \right],  
\end{eqnarray}
where $d = {\mathrm{dim}}({\cal S})$ is the number of signal parameters.
In Ref.\ \cite{Flanagan.E:1998} this relation was used 
to define an effective dimension for any space of signals, and that effective
dimension was used instead of ${\cal N}_{\text{eff}}(Q_0)$ to
characterize the signal space.  Here however we will
instead parameterize our comparisons directly in terms of ${\cal
N}_{\text{eff}}$.

We also note that for this special case of a linear signal subspace,
we have
\begin{equation}
{\cal N}_{\text{eff}}(A_\star) \approx 2^{I_s(A_\star)},
\end{equation}
where $I_s(A_\star)$ is the number of bits of information about the source
carried by a detected signal with signal-to-noise ratio $A_\star$, as
defined in Ref.\ \cite{flanagan.e:1998a}.  We conjecture that this
relation might be approximately valid for general signal manifolds.


We now turn to the relative performance of the 
excess power and matched filtering search methods.
As explained in the previous subsection, once we specify the false
alarm probability $Q_0$ and true detection probability $Q_A$ we can
compute the minimum signal amplitude $A_{\text{min}}$ necessary for
detection via the excess power method, as a function of $Q_0$, $Q_A$
and $V$.  Let us denote that 
value here as $A_{\text{min}}^{\text{EP}}$:
\begin{equation}
A_{\text{min}}^{\text{EP}} = A_{\text{min}}(Q_0,Q_A,V).
\label{e:AEPdef}
\end{equation}
Similarly, we can compute the minimum amplitude necessary for detection
with matched filtering with ${\cal N}_{\text{eff}}$ independent
templates; the result is
\begin{equation}
A_{\text{min}}^{\text{MF}} = A_{\text{min}}(Q_0/{\cal N}_{\text{eff}},Q_A,1/2).
\label{e:AMFdef}
\end{equation}
In other words, one simply uses the same formula with $Q_0$ replaced
by $Q_0/{\cal N}_{\text{eff}}$, and with the number of degrees of freedom
$2V$ being unity.
We define the relative effectiveness $\eta$ of the excess power 
method relative to the bank of filters by
\begin{equation}
\eta(Q_0,Q_A,{\cal N}_{\text{eff}},V) = {A_{\text{min}}^{\text{EP}} /
A_{\text{min}}^{\text{MF}}}.
\label{e:eta}
\end{equation}
The factor by which the event rate for the excess power
method is smaller than that for the bank of matched filters is
$\eta^3$.  The relative effectiveness is plotted in Fig.\ \ref{f:eta}
as a function of $V$ and ${\cal N}_{\text{eff}}$.  For $V \le 100$ we see
that $\eta \ge 0.6$ always, showing that the excess power method
performs almost as well as matched filtering.

When the time-frequency window ${\cal T}$ is not known in advance, one
must search over time frequency windows.   This reduces the efficiency
of the excess power method compared to matched filtering.  An
approximate parameterization of this reduction can be obtained by
replacing in Eq.\ (\ref{e:AEPdef}) the false alarm probability $Q_0$
by $Q_0 / {\cal N}_{\text{w}}$, where ${\cal N}_{\text{w}}$ is
the number of statistically independent time frequency windows
searched over per start time.  We have performed Monte Carlo
simulations with white Gaussian noise which suggest that ${\cal
N}_{\text{w}} \alt 100 \, V_{\text{max}}$, where  
$V_{\text{max}}$ is the largest time-frequency volume searched over.  The
resulting change in the relative efficiency $\eta$ is not very large.

Further insight into the relation between the excess power and
matched filtering methods can be obtained as follows.
First, an approximate formula for the
function (\ref{e:Amindef}) is obtained by approximating the
distribution of ${\cal E}$ to be a Gaussian:
\begin{eqnarray}
&&A_{\text{min}}(Q_0,Q_A,V)^2 = A_\star(Q_0,V)^2 \nonumber \\
\mbox{} && 
\ \ + 2 \sqrt{2} \, {\mathrm{erf}}^{-1}(2 Q_A -1) \sqrt{V + A_\star(Q_0,V)^2},
\label{e:Aminapprox}
\end{eqnarray}
where $A_\star(Q_0,V)$ is obtained by inverting Eq.\
(\ref{e:threshold_approximate}).   This formula is typically accurate
to a few percent for $0.5 \alt Q_A \alt 0.99$.  
When $Q_0 \ll 1 - Q_A$ (which will typically be the case), we can
neglect the second term in Eq.\ (\ref{e:Aminapprox}) in comparison
with the first, so that
\begin{equation}
A_{\text{min}}^{\text{EP}}(Q_0,Q_A,V) \approx A_\star(Q_0,V).
\end{equation}
Second, the quantity $A_{\text{min}}^{\text{MF}}(Q_0,Q_A)$ will
similarly depend weakly on $Q_A$ and will be well approximated by 
the quantity $A_\star(Q_0)$ obtained from Eq.\ (\ref{e:th_ideal2}).
Combining these approximations together with Eq.\
(\ref{e:threshold_approximate}), we see
that the excess power method is equivalent (in terms of detection
thresholds) to matched filtering with an effective number
of templates of 
\begin{equation}
{\cal N}_{\text{eff}} = \left( 1 + {A_\star^2 \over 2 V} \right)^V {\sqrt{V}
\over A_\star} \sim \left( 1 + {A_\star^2 \over 2 V} \right)^{V}.
\label{Ntmax}
\end{equation}
The quantity (\ref{Ntmax}) was shown in Ref.\ \cite{flanagan.e:1998a}
to be the total number of distinguishable signals within the given
time-frequency window with signal-to-noise $\le A_\star$.  
In other words, it is the maximum possible value for the effective
number of independent templates ${\cal N}_{\text{eff}}$ for \emph{any}
manifold of signals inside the time-frequency window $V_{\cal T}$, as a
function of the time-frequency volume $V$.
Hence, we can understand the excess power method as a limiting case of
matched filtering: the case where the manifold of signals 
becomes so large (perhaps curving back and intersecting
itself) that (when smeared out by the noise) it effectively fills up
the entire space $V_{\cal T}$ of signals within the given time-frequency
window.  The equivalence can also be seen from the fact, noted above,
that the excess power statistic coincides with the matched filtering
statistic when dependence of the signal on its parameters is
linear and the number of parameters  
coincides with the dimension $2 V$ of $W_{\cal T}$ \cite{flanagan.e:1998a}.

\section{BAYESIAN ANALYSIS OF SIGNAL DETECTION}
\label{s:bayes_analysis}

In this section we show how our proposed search method arises
naturally from an analysis of the detection of signals from a Bayesian
point of view.  Section \ref{ss:prior} defines the class of signals
under consideration in terms of a prior probability density function
(PDF).   Section \ref{ss:derivation} derives the excess power statistic,
and Sec.\ \ref{ss:thresholds} compares detection criteria based on
a false alarm rate to criteria based on the probability that a signal is
present in the data.

\subsection{The space of signals}
\label{ss:prior}

The signals of interest (e.g., black hole mergers) are poorly understood.  We
characterize our knowledge in terms of a prior PDF
$p({\mathbf{s}})$ for signals ${\mathbf{s}}$ in the vector space
$\mathcal{V}$.  In this subsection we explain how to encode knowledge of the
expected bandwidth and duration of the signals in the PDF.

Suppose we know that the signal ${\mathbf{s}}$ lies approximately
within some time-frequency window ${\cal T} = [t_s,t_s + \delta t]
\times [f_s,f_s + \delta f]$, but that nothing else is known about the
signal.  Then we know that ${\mathbf{s}}$ belongs to a subspace
${\mathcal{V}}_{\cal T}$ of $\mathcal{V}$.  Of course, there are several
slightly inequivalent choices of such a subspace, as discussed in
Sec.\ \ref{s:definition} above, but we will assume that these differences
are unimportant, and pick one choice of ${\cal V}_{\cal T}$.

For any vector ${\mathbf{h}}$ in $\mathcal{V}$, we can write
${\mathbf{h}}={\mathbf{h}}_\parallel+{\mathbf{h}}_\perp$, where
${\mathbf{h}}_\parallel$ is the projection of ${\mathbf{h}}$ into
${\mathcal{V_T}}$ and ${\mathbf{h}}_\perp$ is perpendicular to
${\mathcal{V_T}}$.  We assume the following form for the prior PDF
$p({\bf s} | {\cal T})$ given the time-frequency window ${\cal T}$:
\begin{eqnarray}
p({\bf s} | {\cal T}) d^N {\bf s} &=& 
\delta^{(N-2V)}({\mathbf{s}}_\perp)\,d^{(N-2V)}{\mathbf{s}}_\perp
\nonumber \\
\mbox{} && \times p_1(A) d^{2V} {\bf s}_\parallel,
\label{e:sPDF}
\end{eqnarray}
where $A^2 = ({\bf s}_\parallel,{\bf s}_\parallel)$ and $N$ is the
dimensions of ${\cal V}$.  
Here the first factor consisting of the ($N-2V$)-dimensional $\delta$ function
restricts ${\bf s}$ to lie in ${\cal V}_{\cal T}$.  The second factor
depends only on the magnitude $A$ of ${\bf s}_\parallel$, which
means that we assume all directions in the vector space
${\mathcal{V_T}}$ are equally likely when one measures lengths and angles with
the inner product~(\ref{e:innerProduct}).%
\footnote{It would be more
realistic to make this assumption with respect to an inner product
on ${\cal V}$ whose definition did not depend on the noise spectrum,
but if the noise spectrum does not vary too rapidly
within the bandwidth of interest, the distinction is 
not too important and our assumption will be fairly realistic.}
We can rewrite the prior PDF (\ref{e:sPDF}) as 
\begin{eqnarray}
\label{e:prior} 
  p({\mathbf{s}}|{\cal T})\,d^N{\mathbf{s}} &=&
  \delta^{(N-2V)}({\mathbf{s}}_\perp)\,d^{(N-2V)}{\mathbf{s}}_\perp
  \nonumber \\
  &&\quad \times \frac{\Gamma(V)}{2\pi^V}\,d^{(2V-1)}\Omega_\parallel
\; p(A)\,dA  
\end{eqnarray}
where $d^{(2V-1)}\Omega_\parallel$ is the $(2V-1)$-dimensional element of
solid angle and where $p(A) dA$ is the probability that the signal
amplitude lies between $A$ and $A + dA$.  We discuss the choice of
$p(A)$ in Sec.\ \ref{ss:thresholds} below.

So far we have assumed that the time interval $[t_s,t_s+\delta t]$ and
frequency interval $[f_s,f_s + \delta f]$ are known.  In a real
search, however, one must account for ignorance of these parameters.
An appropriate prior which does this is
\begin{equation}
\label{e:prior3}
p({\mathbf{s}})\,d^N{\mathbf{s}} =
  p({\mathbf{s}}|{\cal T})\,d^N{\mathbf{s}}\; p_{\cal T}({\cal T}) d^4
  {\cal T},
\end{equation}
where $p_{\cal T}({\cal T}) = p_{\cal T}(t_s,\delta t,f_s,\delta f)$
is a prior PDF on the time-frequency window parameters.  
The PDF $p_{\cal T}({\cal T})$ should be uniform in $t_s$, but its
dependence on the parameters $\delta t$, $f_s$ and $\delta f$ will 
depend on the class of sources under consideration.  We will see below
that our analysis depends only weakly on the choice of PDF $p_{\cal
T}({\cal T})$, as long as it is a slowly varying function of its
parameters. 

\subsection{Derivation of the search method}
\label{ss:derivation}

In the Bayesian approach to signal detection there is a unique and optimal
method to search the data stream for signals if the statistical properties of
the detector noise and the prior probability distribution for signals are
known.  One computes the probability $p_s({\mathbf{h}})$ that some
signal ${\mathbf{s}}$ is present in the measured data $\mathbf{h}$.  The
signal-detection criterion is that the probability
$p_s({\mathbf{h}})$ exceeds some threshold value.  This is the
starting point for our analysis; more details can be found in Wainstein and
Zubakov~\cite{wainstein.l:1962} or Finn and
Chernoff~\cite{finn.l:1992,finn.l:1993}.

The prior PDF given Eqs.~(\ref{e:prior}) and (\ref{e:prior3})
describes our state of knowledge about the signals to be searched for.  
Now let $p_{s0}$ denote the a priori probability that
gravitational waves exist (or that our signal model is correct), for
which an appropriate value for the first searches might be $p_{s0} =
1/2$.  The signal PDF (\ref{e:prior3}) then gets modified to
\begin{equation}
(1 - p_{s0}) \delta^N({\bf s}) \, d^N {\bf s} + p_{s0} \, p({\bf s})
\, d^N {\bf 
s}.
\end{equation}
It then follows that the posterior probability $p_s({\bf
h})$ that a signal is present in the data $\mathbf{h}$ is given by
\begin{equation}
  \label{e:probOfSignal}
{p_s({\bf h}) \over 1 - p_s({\bf h})} = \Lambda({\bf h}) { p_{s0} \over 1
- p_{s0}},
\end{equation}
where the likelihood function $\Lambda({\mathbf{h}})$ is 
\begin{eqnarray}
  \label{e:likelihood}
  \Lambda({\mathbf{h}}) &=& \int
  \Lambda({\mathbf{h}};{\mathbf{s}})\,
  p({\mathbf{s}})\,d^N{\mathbf{s}} \nonumber \\
  &=& \int\!\!\!\!\int \Lambda({\mathbf{h}};{\mathbf{s}})\,
  p({\mathbf{s}}|{\cal T})p({\cal T})\,d^N{\mathbf{s}}\,d^4{\cal T}
\end{eqnarray}
and
\begin{equation} 
 \Lambda({\mathbf{h}};{\mathbf{s}}) =
  \frac{p({\mathbf{h}}|{\mathbf{s}})}{p({\mathbf{h}}|{\mathbf{s}}={\mathbf{0}})}
  .  \label{e:Lambdadef}
\end{equation}
In Eq.\ (\ref{e:Lambdadef}) the quantity $p({\mathbf{h}}|{\mathbf{s}})$ is the
probability of measuring the time series ${\mathbf{h}}$ when the
signal ${\mathbf{s}}$ is present, and 
$p({\mathbf{h}}|{\mathbf{s}}={\mathbf{0}})$ is the corresponding probability
when no signal is present.   For stationary Gaussian noise
the likelihood ratio
$\Lambda({\mathbf{h}};{\mathbf{s}})$ is~\cite{finn.l:1992}.
\begin{equation}
  \label{e:likelihood;s}
\Lambda({\mathbf{h}};{\mathbf{s}}) =
  \exp[({\mathbf{h}},{\mathbf{s}})-({\mathbf{s}},{\mathbf{s}})/2] .
\end{equation}
Equation~(\ref{e:probOfSignal}) shows that the probability
$p_s({\mathbf{h}})$ increases monotonically with increasing
$\Lambda({\mathbf{h}})$.  Consequently,  thresholding on
$\Lambda({\mathbf{h}})$ to detect signals is equivalent to thresholding on the
probability that a signal is present in the data stream.  This is also the
optimal signal detection strategy in the Neyman-Pearson sense of maximizing
the detection probability for a fixed false alarm
probability~\cite{wainstein.l:1962,Kassam.S:1988}.

The integral (\ref{e:likelihood}) includes a integral over all
possible time-frequency windows ${\cal T}$, which can be approximated
as a sum:
\begin{equation}
  \Lambda({\mathbf{h}}) \approx {1 \over {\cal N}_{\text{windows}} }
  \sum_{\cal T} \ \int
  \Lambda({\mathbf{h}};{\mathbf{s}})\,
  p({\mathbf{s}}| {\cal T})\,d^N{\mathbf{s}}.
\label{e:likelihood32}
\end{equation}
Here ${\cal N}_{\text{windows}}$ is the number of grid points in a grid
on the four dimensional space of time-frequency windows used to
approximate the integral.  Now if a signal is present, the summand in
Eq.\ (\ref{e:likelihood33}) will be a sharply peaked function of
${\cal T}$.  If the grid spacing is chosen to approximately coincide
with the width of the peak (so that ${\cal N}_{\text{windows}}$ is the
number of statistically independent time-frequency windows)
then the sum will be dominated by the largest term, and we obtain
\begin{equation}
  \Lambda({\mathbf{h}}) \approx {1 \over {\cal N}_{\text{windows}} }
  \max_{\cal T} \ \int
  \Lambda({\mathbf{h}};{\mathbf{s}})\,
  p({\mathbf{s}}| {\cal T})\,d^N{\mathbf{s}}.
\label{e:likelihood33}
\end{equation}
Thus it is sufficient to consider only a single
time-frequency region in the remainder of this section with the understanding
that the signal detection will be based on the maximum of the likelihood
function over all relevant time-frequency windows.  
Also we can factor ${\cal N}_{\text{windows}}$ as ${\cal N}_{\text{windows}} =
{\cal N}_{\text{st}} \, {\cal N}_{\text{w}}$, where ${\cal N}_{\text{st}}$ is
the number of statistically independent starting times $t_s$ in the
search, and ${\cal N}_{\text{w}}$ is the number of statistically
independent time-frequency windows per start time.

The evaluation of the integral in Eq.~(\ref{e:likelihood33}) with the
prior PDF given 
by Eq.~(\ref{e:prior}) can be done in stages.  First we integrate over the
delta-function to restrict the possible signals to the vector space
${\mathcal{V}}_{\cal T}$.  This essentially replaces ${\mathbf{s}}$ by
${\mathbf{s}}_\parallel$ in Eq.~(\ref{e:likelihood;s}).  
Next we use the definition (\ref{e:calEdef1}) of ${\cal E}$
to write the inner product appearing in Eq.\ (\ref{e:likelihood;s}) as
\begin{equation}
({\mathbf{h}},{\mathbf{s}}_\parallel)=A{\mathcal{E}}^{1/2}\cos\theta,
\end{equation}
where $\theta$ is the angle between the vectors ${\mathbf{h}}_\parallel$ and
${\mathbf{s}}_\parallel$.  We then obtain the formula
\begin{equation}
\label{e:likelihoodFunction}
  \Lambda({\mathbf{h}}) = {1 \over {\cal N}_{\text{st}} \,
  {\cal N}_{\text{w}}} \, \int \Lambda({\mathbf{h}};A)\,p(A)\,dA
\end{equation}
with
\begin{eqnarray}
\label{e:likelihood;A}
  \Lambda({\mathbf{h}};A) &=&
  \frac{\Gamma(V) \, e^{-A^2/2}}{\pi^{1/2}\Gamma(V-1/2)}
  \int_0^\pi e^{A{\mathcal{E}}^{1/2}\cos\theta}\sin^{2V-2}\theta\,d\theta
  \nonumber\\
  &=& \Gamma(V)e^{-A^2/2}
  (A{\mathcal{E}}^{1/2}/2)^{1-V} I_{V-1}(A{\mathcal{E}}^{1/2}) \nonumber\\
  &=& p({\mathcal{E}}|A,V)/p({\mathcal{E}}|A=0,V)
\end{eqnarray}
[cf.\@ Eq.~(\ref{e:excessPowerProbabilityDistribution1})].
Here $I_n(x)$ is the modified Bessel function of the first kind of order $n$,
and maximization over time frequency windows ${\cal T}$ is understood.

The quantity (\ref{e:likelihood;A}) is a monotonically increasing
function of the 
power $\mathcal{E}$.  Hence thresholding on $\Lambda$ is equivalent to
thresholding on ${\cal E}$, and so ${\cal E}$ is the optimal (in
the Bayesian sense) statistic for the detection of the class of
signals we have considered.

\subsection{Bayesian thresholds}
\label{ss:thresholds}

Frequentist detection thresholds ${\cal E}^\star$ are set by specifying a
false alarm 
rate, and can be computed using Eq.~(\ref{e:falseAlarm}).  As is well
known, if such a threshold is exceeded it does not necessarily
mean that a signal is present with high probability, even
for low false alarm probabilities
\cite{Lindley_L:1957,Shafer_G:1982,Finn_L:1997}.  To determine how
likely it is that a signal is actually present in the data stream
requires the use of Bayesian methods. 

For a Bayesian detection strategy, one sets a threshold on the posterior
probability $p_s({\bf h})$ that a signal is present given the data.
This probability is related to the likelihood function $\Lambda({\bf
h})$ by Eq.\ (\ref{e:probOfSignal}).  In general, the integral
(\ref{e:likelihoodFunction}) required to compute
$\Lambda({\mathbf{h}})$ must be performed numerically.  Since
$\Lambda({\bf h})$ depends on the data ${\bf h}$ only through ${\cal
E}({\bf h})$, one can determine a Bayesian threshold for ${\cal E}$
from the value of $p_s$.

Consider a search characterized by ${\cal N}_{\text{st}}$ statistically
independent start 
times and ${\cal N}_{\text{w}}$ statistically independent time-frequency
windows.  The frequentist false alarm probability $Q_0$ of the
previous section [Eq.\ (\ref{e:falseAlarm})] 
is the false alarm probability for a given start
time and a given time-frequency window.  Hence the false alarm probability
for the entire search is
\begin{equation}
p_{\text{fa}}({\cal E}^\star) = Q_0({\cal E}^\star) {\cal N}_{\text{st}}
{\cal N}_{\text{w}}.
\label{e:pfadef}
\end{equation}
It is natural, in comparing frequentist and Bayesian thresholds, to
set $p_s = 1 - p_{\text{fa}}$.  For example, for ``$99\%$ confidence''
one would choose $p_s = 0.99 = 1 - p_{\text{fa}}$.  We emphasize that this means
``$99\%$ confidence that events will be due to signals'' for the Bayesian,
while it means ``$99\%$ confidence that there will be no false events'' for
the frequentist; since these are different statistical statements, the
frequentist and the Bayesian will obtain different thresholds.

We first discuss approximate evaluation of Bayesian thresholds.
The integral (\ref{e:likelihoodFunction}) can be approximately
evaluated in the the regime $V \gg 1$ by using the Laplace
approximation, if the prior PDF $p(A)$ does not vary too rapidly.
The result is
\begin{eqnarray}
\Lambda({\bf h}) &=& { \sqrt{2 \pi V} \over {\hat A}} \, {p({\hat A})
\over {\cal N}_{\text{st}} {\cal N}_{\text{w}}}
\, \left( 1 + {{\hat A}^2  \over 2 V} \right)^{-V} 
\exp({\hat A}^2/2) \nonumber \\
\mbox{} && \times \left[ 1 + O\left({1 \over V} \right) 
+ O\left({1 \over {\hat A}^2} \right) 
+ O\left({V \over {\hat A}^4} \right) \right] \; ,
\label{Lambda14}
\end{eqnarray}
where ${\hat A} = {\hat A}({\bf h})$ is defined by
\begin{equation}
{\cal E}({\bf h}) =   2 V + {\hat A}({\bf h})^2.
\end{equation}
If we now compare Eqs.\ (\ref{e:threshold_approximate}),
 (\ref{e:probOfSignal}), (\ref{e:pfadef}) and (\ref{Lambda14}) and use
 $p_{s0}=1/2$ and $1-p_s \ll 1$, 
we see that the Bayesian threshold ${\hat A}$ and frequentist
threshold $A_\star$ are related by
\begin{equation}
\left( 1 + {{\hat A}^2 \over 2 V} \right)^{-V} \, e^{ {\hat A}^2/2} =
{\cal F} \ \left( 1 + {A_\star^2 \over 2 V} \right)^{-V} \, e^{ A_\star^2/2},
\label{e:comparethresholds}
\end{equation}
where the factor ${\cal F}$ is 
\begin{equation}
{\cal F} =  { {\hat A} A_\star^2 \over 2 V p({\hat A}) }.
\end{equation}
Clearly the two thresholds coincide when ${\cal F}=1$.  However,
typically the factor ${\cal F}$ can be quite significant and can cause
the Bayesian and frequentist thresholds to differ substantially.

It is useful to consider a specific example.  Suppose that we are
searching for black hole mergers which we expect to produce short (a few ms)
broad band signatures with a time-frequency volume of $V=100$.  We
want to be $99\%$ sure of our detection, so we set 
$p_s=0.99 = 1 - p_{\text{fa}}$.  Suppose that the search duration is 1/3 of
a year, so that the number of independent start time is ${\cal N}_{\text{st}}
= 10^{10}$ say, and that the number of statistically
independent windows is $10^4$.  For $99\%$ confidence that there
will be no false alarms, we should choose $Q_0 = 10^{-16}$, from Eq.\
(\ref{e:pfadef}).  This gives from Eq.\ (\ref{e:falseAlarm})
a frequentist threshold of ${\cal E}^\star = 411.3$ corresponding to a
signal-to-noise threshold of $A_\star = 14.5$.  The corresponding Bayesian
threshold depends on the specification of prior PDF for signal
amplitudes $A$.  A reasonable choice of prior is 
\begin{equation}
\label{noisespec}
p(A) = \left\{ \begin{array}{ll} {3 A_c^3 / A^4} 
        & \mbox{$A \ge A_c $,}\\
    0 & \mbox{$A < A_c$.}\\
\end{array} \right.
\end{equation}
This is just the distribution that would be expected for sources
distributed uniformly in time and space, except that it is cutoff in
an approximate way at small $A$ in order to ensure correct
normalization.  The parameter $A_c^3$ is prior probability per
start-time of an event being present with signal-to-noise ratio
exceeding unity.   
Based on population estimates such Ref.~\cite{Zwart.S:1999}, we
optimistically assume a prior probability of order unity for
approximately one merger event per year with $A>1$, which translates into
$A_c^3 \sim 10^{-10}$.  We can now compute a Bayesian threshold by 
combining Eqs.\ (\ref{e:probOfSignal}), (\ref{e:likelihoodFunction}),
and (\ref{e:likelihood;A}).  The result is ${\cal E} = 539$
corresponding a signal-to-noise ratio threshold of ${\hat A} = 18.4$,
which is substantially higher than the frequentist value of $14.5$.

\section{Implementation}
\label{s:computation}

As discussed in Sec.\ \ref{s:definition}, one will not know in advance
the start time $t_s$, duration $\delta t$ and 
frequency band $[f_s, f_s + \delta f]$ of signals in a real search,
and thus one must 
perform a search over these four parameters.  One needs to compute the
excess power ${\cal E}_{\cal T}({\bf h})$ for each possible
time-frequency window, and record as possible events all of those
windows for which ${\cal E}_{\cal T}$ is above threshold.%
\footnote{Note that different thresholds will be required for each
window ${\cal T}$, but the false alarm probability $Q_0$ will be the
same for each window.}
We assume that we wish to search over all
values of $\delta t$ in the range 
\begin{equation}
\delta t_{\text{min}} \le \delta t \le \delta t_{\text{max}},
\label{e:timerange}
\end{equation}
and over all $f_s$ and $\delta f$ with 
\begin{equation}
 \left. \begin{array}{lll} 
 \delta f_{\text{min}} & \le &\delta f \le \delta f_{\text{max}} \\
  f_{\text{min}} &\le& f_s  \\
  f_s + \delta f &\le& f_{\text{min}} + \delta f_{\text{max}} \\
\end{array} \right\}.
\label{e:freqrange}
\end{equation}
It is clear that the computational cost
can quickly grow to unreasonable proportions, so it is important to
achieve an efficient implementation of the search technique.

There are (at least) two different ways to implement a search over a
pre-specified set of time-frequency windows.  The first
uses many FFTs of data segments with durations in the range
(\ref{e:timerange}) as suggested by the derivation in
Sec.~\ref{s:detection}, and for each FFT computes $\mathcal{E}$ for all
frequency bands in the range (\ref{e:freqrange}).  This
process is then repeated for every possible start time.  We call this
procedure the \emph{short FFT method}.  The second method partitions
the time series into long data segments each containing $M$ samples, 
and for each of these segments computes its FFT\@.  That FFT is then
partitioned into $\delta f_{\text{max}} / \delta f_{\text{min}}$
non-overlapping frequency bands each of width $\delta f_{\text{min}}$,
and for each one the FFT is bandpass filtered to that
frequency band and then inverse Fourier transformed.  The result is 
$\delta f_{\text{max}}/\delta f_{\text{min}}$ different timeseries,
which we call channels, each containing particular frequency
information.   The elements of these time series are then squared.
One obtains in this way a time-frequency plane in which each pixel
represents the total power in a time-frequency volume of order
$\sim1$.  Finally, one computes the total power in the various rectangles
in this time-frequency plane.  We call this procedure the \emph{long FFT 
method}.  We now consider the computational cost of each method in
turn and argue that the second method is more computationally efficient.

\subsection{Sufficiency of approximate version of statistic}
\label{sss:FourierBasis}

In Sec.\ \ref{s:definition} above we discussed how to compute an
approximate version of the excess power statistic [cf.\ Eq.\
(\ref{e:calEapprox})].  Namely, for a given start
time $t_s$, perform a Fourier transform of the $K$ point time series
$\{h_j\}$ corresponding to the time window $\delta t$, where $K = \delta t /
\Delta t$.  Denote the discrete Fourier Transform (DFT) by 
$\tilde{h}_k$ where $0\leq k \leq K/2$.  Identify the frequency
components $k_1 \le k < k_2$ of $\tilde{h}_k$ which belong to the frequency
band $\delta f$, and construct the statistic
\begin{equation}
  {\mathcal{E}} = 4{\textstyler{\sum_{k_1 \le k < k_2}}} |\tilde{h}_k|^2/S_k,
\label{e:excessPower2}
\end{equation}
where $S_k$ is the noise power spectrum defined in Eq.\ (\ref{e:spectrumdef}).

The quantity (\ref{e:excessPower2}) differs from the exact statistic
${\cal E}$ due to the fact that the expectation value $\langle
\tilde{h}_k \tilde{h}_{k'}^* \rangle$ is not diagonal.  (It becomes
effectively diagonal only in the limit $\delta t \to \infty$.)  
Consequently, the expression (\ref{e:excessPower2}) is not a sum of
squares of independent unit-variance Gaussian random variables, and
so its distribution could in principle differ from the non-central
$\chi^2$ distribution.  However, in practice, if the power spectrum of
the noise is a slowly varying function of frequency, then 
the correlations introduced by using the expression
(\ref{e:excessPower2}) are small.  To confirm this, we  
have examined the behavior of the statistic (\ref{e:excessPower2})
computed from the DFT of colored, Gaussian noise.  We generated
colored noise according to the correlation generating scheme 
\begin{equation}
n_j = \big(m_j - 0.8 \, m_{j-1} + 1.2944 \, n_{j-1}
 - 0.64 \, n_{j-2}\big) / 1.3145
\end{equation}
where $m_j$ are uncorrelated Gaussian deviates and
$n_j=m_j=0$ for $j<0$. To determine detection statistics, we used the signal
model 
\begin{equation}
s_j = S \exp[ -16(j/2N-1)^2 ] \cos(2\pi jf_0)
\end{equation}
with  $N=4096$ samples in the signal.   The central frequency of the
signal was $f_0=600/4096$ and 
the constant $S$ was chosen to give the required value of signal
amplitude $A$.  We found the operating characteristics of 
$\mathcal{E}$ were not significantly affected 
by using the approximate formula (\ref{e:excessPower2}) rather than
the exact formula.  This is demonstrated in Figs.~\ref{f:falseAlarm}
and \ref{f:trueDetect} where we have overlaid simulated false alarm and
true detection probabilities on top of the distributions computed in
Sec.~\ref{ss:operation}.  We calculated the goodness of fit using a
$\chi^2$ test for a few of the curves in these figures.  In each case, the
reduced $\chi^2$ value was $\leq 1.03$, indicating that it is unlikely that
the simulated data is drawn from distributions other than those presented in
Sec~\ref{ss:operation}.  We therefore conclude that we can use the 
approximate formula (\ref{e:excessPower2}) without significantly
modifying the behavior of the statistic $\mathcal{E}$.

\subsection{The short FFT method}
\label{ss:shortFFT}

The algorithm is: (1) pick a start time, (2) pick a time duration $\delta t$,
(3) FFT the selected data and compute the power in each frequency bin, (4) sum
the power in the bands of interest, (5) loop over steps (2)--(4) until all time
durations are used, and (6) repeat steps (1)--(5) for all start times.

The computational cost for steps (3)--(4) can be estimated as follows.  The
number of data points in segment of data of duration $\delta t$ is
$N=\delta t/\Delta t$, so 
each FFT requires $3N\log_2N$ floating point operations.  Since the
relevant frequency band has a dimensionless bandwidth%
\footnote{By dimensionless bandwidth we mean number of frequency bins,
i.e., bandwidth multiplied by $\Delta t$.  Note that the dimensionless
  bandwidth of the entire frequency 
  band up to the Nyquist frequency is $1/2$.}
$\alpha_{\text{max}}=\delta f_{\text{max}}\Delta t$, the total cost to compute
the power in each frequency bin in the band is
  ${3}N\alpha_{\text{max}}$ operations.
Now, for the $k$th frequency bin, it costs $(N\alpha_{\text{max}}- k)$
floating point operations to compute the power in all frequency intervals
whose have lowest frequency component is in the $k$th bin; the number
of operations required to do this for all frequencies $k$ is
\begin{equation}
  \sum_{k=1}^{N\alpha_{\text{max}}- 1} (N\alpha_{\text{max}} - k) = 
  \frac{N\alpha_{\text{max}}}{2} 
  \bigl(N\alpha_{\text{max}}- 1\bigr).
\end{equation}

These steps must be repeated for each $N$ with $N_{\text{min}} \le N \le
N_{\text{max}}$, where $N_{\text{min}}=\delta t_{\text{min}}/\Delta t$ and
$N_{\text{max}}=\delta t_{\text{max}}/\Delta t$.  Thus, the total
computational cost per start time $C_{\text{short}}$ is
\begin{equation}
C_{\text{short}} = \sum_{N=N_{\text{min}}}^{N_{\text{max}}} 
    N[ 3\log_2N + {3}\alpha_{\text{max}}
    + \case{1}{2}\alpha_{\text{max}}(N\alpha_{\text{max}}-1) ].
\end{equation}
One will typically have $N_{\text{min}} \sim 1$ and
$N_{\text{max}}\alpha_{\text{max}}=V\gg1$, where $V$ is the total
time-frequency volume to be searched. In this case, a useful approximation to
the computational cost per start time is
\begin{equation}
\label{e:shortCost}
  C_{\text{short}} \simeq \frac{V^2}{2\alpha_{\text{max}}} \left(
    \frac{3\log_2V}{\alpha_{\text{max}}} + \frac{V}{3} \right).
\end{equation}
The total computational cost in flops (floating-point operations per
second) can be obtained by multiplying $C_{\text{short}}$ by the
sampling rate.

\subsection{The long FFT method}
\label{ss:longFFT}

As discussed above, in the long FFT method one constructs a
time-frequency plane consisting of $N_{\text{channels}} = 
\delta f_{\text{max}}/\delta f_{\text{min}}$ different channels.
The power in any time-frequency window can then be computed by summing
the power in that region of the time-frequency plane.  
The data stream is broken into chunks of length $M$ points, each chunk
is FFTed, and the requisite number of channels are produced by
bandpass filtering, Fourier transforming back into the time domain,
and squaring the time samples.  For each chunk, the computational cost
of this step is
\begin{equation}
  C_1 = M[3 (1 + N_{\text{channels}})
  \log_2 M + N_{\text{channels}}].
\end{equation}

To search over the time-frequency plane,   we first pick the frequency
interval $\delta f$ and construct the $\delta f_{\text{max}}/\delta f$
channels of this bandwidth; this requires $M(\delta f/f_{\text{min}}-1)$
additions.  For each of these new channels,   we sum up the power in various
time intervals.  This step requires $N_{\text{max}}$ operations per start
time, of which there are $M-N_{\text{min}}$.  Thus the total cost at this
stage of the search is 
\begin{equation}
  C_2 = \sum_{j = 1}^{N_{\text{channels}}}
  \frac{N_{\text{channels}}}{j}
  [ M (j - 1) + N_{\text{max}}(M-N_{\text{min}}) ].
\end{equation}
Since there are approximately $M$ different start times, the cost per start
time is given by the approximate formula
\begin{equation}
\label{e:longCost}
  C_{\text{long}}\simeq\alpha_{\text{max}}^{-1}V^2\ln V.
\end{equation}

\subsection{Comparison of the two methods}
\label{ss:comparison}

The space of time-frequency windows to search over was delineated in
the Introduction for the initial interferometers in LIGO\@.  We
adopt the corresponding parameter values $\delta
f_{\text{min}}=2\>\text{Hz}$, 
$\delta f_{\text{max}}=200\>\text{Hz}$,
$\delta t_{\text{min}}=0.005\>\text{s}$, and
$\delta t_{\text{max}}=0.5\>\text{s}$.
The computational power required using the long FFT method is
$0.3\>\text{GFlops}$,  which saves a factor of $\sim 14$ over the
short FFT method if $\Delta t = 0.001$ seconds. 

In general, the computational gain afforded by the long FFT method over the
short FFT method is given approximately by 
\begin{equation}
  \frac{C_{\text{short}}}{C_{\text{long}}}
    \sim \frac{3}{2\ln2}\frac{1}{\alpha_{\text{max}}}
    + \frac{1}{6}\frac{V}{\ln V}.
\end{equation}
The first term shows that there is at least a factor of $\sim4$ to be gained
by the long FFT method; in addition, the computational gain increases
with the total time-frequency volume $V$.  For $V=100$, the second
term is also $\sim4$. 

There is a further benefit to the long FFT technique.  
It allows finer frequency resolution in the choice of starts $f_s$ and
ends $f_s + \delta f_s$ of the frequency bands to be explored
(although the above estimates of computational cost were for a search
equivalent to the short FFT search).  Moreover, as part of a
hierarchical search, 
the long FFT method has a further advantage in that it allows follow
ups to be made without significant further computations.  The next
stage of a hierarchical search might involve techniques other than the
excess-power method, e.g., Hough transforms or other line tracking
algorithms.

\section{Multiple detectors}
\label{s:multipleDetectors}

The network of gravitational wave detectors under construction around the
world brings benefits that a single instrument cannot.  This is especially
true for ``blind'' search techniques, such as the power statistic.  Since
these techniques do not require the signal to have a specific form, random
noise glitches are much more likely to meet the detection criteria
than is the case for signal-specific searches such as matched
filtering.  Multiple-detector   
statistics will be much more efficient at rejecting such false alarms than
single-detector statistics \cite{creighton.j:1999a,creighton.j:1999b}.
In this section we consider the construction of the optimal detection
strategy for a 
network of detectors.  The derivation requires further formal development.
For maximum clarity, we introduce most of our notation in
Sec.~\ref{ss:multiNotation}.  We derive the multi-instrument detection
statistic for a network of
aligned detectors in Sec.~\ref{ss:multiAligned}.  The two LIGO interferometers
at the Hanford site form such a network.  In addition, if we ignore
the slight misalignment that arises from curvature of the earth, we
can also include the interferometer in Livingston to form a three
interferometer network.  The general case when not all instruments are
aligned is treated in Sec.~\ref{ss:multiGeneral}.

Our analysis is based on the formalism of
Ref.\ \cite{flanagan.e:1998a} which followed earlier work of Ref.\
\cite{Dhurandhar.S:1988}.  We assume that the noise of the detector
network is Gaussian.  Even though we allow correlations 
between noise in different instruments, the assumption of Gaussian noise is a
serious limitation since the main benefit of having several detectors is to
combat non-Gaussian noise.  It should be possible to adapt the theoretical
models of non-Gaussian noise given in Ref.~\cite{creighton.j:1999b} in order
to derive robust multi-detector statistics.  However, it is necessary to
understand first the Gaussian case.

\subsection{Notation and terminology}
\label{ss:multiNotation}

Suppose the detector network consists of $n_d$ detectors.  Denote the output
of the entire network by the vector of time series
\begin{eqnarray}
\vec{\mathbf{h}} &= \{& h_j^A \} \nonumber\\
&= \{& h^1(0), h^1(\Delta t),\ldots,h^1[(N-1)\Delta t],\nonumber\\
&& h^2(0),\ldots,h^2[(N-1)\Delta t],\ldots,h^{n_d}[(N-1)\Delta t]\}
\end{eqnarray}
where $A \in \{1, 2, \ldots, n_d\}$ and $j \in \{0,1,\ldots, N-1\}$.
We assume that the noise $\vec{\mathbf{n}}$ of the network follows a
multivariate Gaussian 
distribution which is determined by the $(N n_d)^2$ correlations
\begin{equation}
\langle n^A(i \Delta t) n^B(j \Delta t) \rangle = C^{AB}(|i-j|\Delta t) =
R_{ij}^{AB} \label{e:multi-correlation}
\end{equation}
where $\langle \cdot \rangle$ denotes an ensemble average.  In general,  the
elements of the network correlation matrix are 
\begin{equation}
({\mathbf{R}})_{A N+i, B N+j} = R^{AB}_{ij}  .
\label{e:RR}
\end{equation}
The convention (\ref{e:RR}) for combining the capital and lower case
indices to form in an $N n_d \times N n_d$ matrix is used from here on.
The probability density function of the noise is given by 
\begin{equation}
p(\vec{\mathbf{n}}) = [(2\pi)^{N n_d}\det{\mathbf{R}}]^{-1/2}
\exp[ -\case{1}{2} (\vec{\mathbf{n}},\vec{\mathbf{n}})]
\end{equation}
where the inner product is given by
\begin{equation}
(\vec{\mathbf{p}}, \vec{\mathbf{q}}) = \sum_{A,B=1}^{n_d}
\sum_{i,j=0}^{N-1} p^A_i Q^{AB}_{ij} q^B_j \label{e:multiInnerProd1}
\end{equation}
and
\begin{equation}
Q^{AB}_{ij} = ({\mathbf{R}}^{-1})_{AN +i, BN +j}  .
\end{equation}
For later convenience, we note that the inner product
(\ref{e:multiInnerProd1}) can be written in terms of the discrete Fourier
transforms of the detector time series, that is
\begin{equation}
(\vec{\mathbf{p}}, \vec{\mathbf{q}}) \simeq 
\sum_{A,B=1}^{n_d} 4 {\mathrm{Re}} \sum_{k=0}^{\lfloor N/2\rfloor}
\tilde{p}^{A\,*}_{k}
({\mathbf{S}}^{-1}_k)^{AB} \tilde{q}^B_{k} 
\end{equation}
where
\begin{equation}
{\mathbf{S}}_k^{AB} \delta_{kk'}
\simeq 2 \sum_{j,j'=0}^{N-1} e^{2 \pi i j k/N} R_{jj'}^{AB} e^{-2
\pi i j' k'/N}  .
\end{equation}
The notation $\lfloor N/2\rfloor$ denotes the greatest integer less than or
equal to $N/2$.
These relations are strictly speaking valid only in the continuum and
infinite time limits $\Delta t\rightarrow0$ and $N \Delta t \rightarrow
\infty$.  Nevertheless, they are sufficiently accurate for most
practical applications.  Finally,  we note that the likelihood
ratio $\Lambda(\vec{\mathbf{h}};\vec{\mathbf{s}})$ is given by
\begin{equation}
  \Lambda(\vec{\mathbf{h}};\vec{\mathbf{s}}) =
  \frac{p(\vec{\mathbf{h}}|\vec{\mathbf{s}})}%
    {p(\vec{\mathbf{h}}|\vec{\mathbf{0}})}
  = \exp[ (\vec{\mathbf{h}},\vec{\mathbf{s}}) - \case{1}{2}
    (\vec{\mathbf{s}}, \vec{\mathbf{s}}) ].
  \label{e:multiLikelihood}
\end{equation}

\subsection{Aligned detectors}
\label{ss:multiAligned}

The simplest type of multi-instrument network to analyze is a network
consisting of instruments which all respond to the same polarization
component of the gravitational wave field.  The two LIGO
interferometers in Hanford form such a detector, and if we ignore the
slight misalignment arising from the curvature of the earth (a $\sim
10\%$ correction effect; see Table \protect{\ref{t:detectors}}) the third
LIGO interferometer in Livingston can also be included.

The signal at any detector is simply a time-delayed version of the
signal that would be detected at the coordinate origin, which for
simplicity we take to be at the center of the earth.  Thus, the signal at
detector $A$ is 
\begin{equation}
s^A(t) = s(t + \tau_A),
\end{equation}
where $\tau_A$ is the time of flight for a gravitational wave between detector
$A$ and the coordinate origin, and $s(t)$ is the signal at the
coordinate origin.  The time delays $\tau_A$ depend on the direction to
the source:  if $\mathbf{m}$ is a unit three vector in the direction of
propagation of the gravitational wave (i.e., opposite to the direction to the
source) and ${\mathbf{x}}^A$ is the location of detector $A$,  then 
\begin{equation}
\tau_A = {\mathbf{m}}\cdot {\mathbf{x}}^A /c
\end{equation}
where $c$ is the speed of light.  Finally, the DFT of the signal at detector
$A$ can be written as 
\begin{equation}
\tilde{s}^A_k = e^{2 \pi i k \Delta_A({\mathbf{m}}) /N} \tilde{s}_k 
\end{equation}
where $\Delta_A({\mathbf{m}}) = \tau_A/\Delta t$ and $\tilde{s}_k$ is the DFT
of the signal at the origin of coordinates.

A convenient description of the multi-instrument network response is the
effective strain $\tilde{h}^{\text{(eff)}}_k$ defined by
\cite{flanagan.e:1998a} 
\begin{equation}
  \tilde{h}^{\text{(eff)}}_k = S^{\text{(eff)}}_k
  {\textstyler{\sum_{A,B=1}^{n_d}}}
	e^{-2 \pi i k \Delta_A({\mathbf{m}}) /N}
  ({\mathbf{S}}^{-1}_k)^{AB} \tilde{h}^B_k 
\end{equation}
where 
\begin{equation}
  1/S^{\text{(eff)}}_k = {\textstyler{\sum_{A,B=1}^{n_d}}}
  e^{-2 \pi i k \Delta_A({\mathbf{m}}) /N} 
	({\mathbf{S}}^{-1}_k)^{AB} e^{2 \pi i k
         \Delta_B({\mathbf{m}}) /N} .
\end{equation} 
We note that the effective strain depends on the direction $\mathbf{m}$ to the
putative source through the time delays $\Delta_A({\mathbf{m}})$.  Generally
the same is true for $S^{\text{(eff)}}_k$, although not
if there are no correlations between the instrumental noise at
separated sites \cite{flanagan.e:1998a}.%
\footnote{It might be reasonable to make this assumption for
  the three detectors at LIGO for example.}
We can now write the likelihood ratio (\ref{e:multiLikelihood}) as
\begin{equation}
\Lambda(\vec{\mathbf{h}};{\mathbf{s}}) 
= \exp[ \lbrak \tilde{h}^{\text{(eff)}}, \tilde{s} \rbrak - \case{1}{2}
\lbrak \tilde{s},\tilde{s} \rbrak ] .
\end{equation}
where 
\begin{equation}
\lbrak \tilde{p},\tilde{q} \rbrak =
4 {\mathrm{Re}} \textstyler{\sum_{k=0}^{\lfloor N/2\rfloor}} \tilde{p}_k
\tilde{q}^*_k/S^{\text{(eff)}}_k .
\end{equation}
The posterior probability $p_s({\vec{\mathbf{h}}})$ that a signal is
present given the data $\vec{\mathbf{h}}$ is determined by integrating the
likelihood against prior probabilities densities for the signal and 
for the source direction ${\bf m}$.  Thus
\begin{equation}
{  p_s(\vec{\mathbf{h}}) 
\over 1 - p_s(\vec{\mathbf{h}}) }
= \Lambda(\vec{\mathbf{h}}) \, { p_{s0} \over 1 - p_{s0} }
\label{e:multiLikelihood2}
\end{equation}
where
\begin{equation}
\Lambda(\vec{\mathbf{h}}) = \int\!\!\!\!\int p(\theta,\phi)d^2 \Omega
\int p({\cal T})d{\cal T} \int p({\mathbf{s}}|{\cal T}) d^N{\mathbf{s}}\,
\Lambda(\vec{\mathbf{h}};\vec{\mathbf{s}})
\label{e:multiLikelihood3}
\end{equation}
and $p_{s0}$ is the a priori probability that any gravitational
wave sources exist.
The mechanism of how information about the
signals is encoded in the prior 
probabilities $p({\mathbf{s}}|{\cal T})$ and $p({\cal T})$ is treated in
Sec.~\ref{ss:prior}, and applies directly to the current context with
only one modification: the inner product $(\cdot,\cdot)$ should be
replaced by $\lbrak\cdot ,\cdot\rbrak$.  In particular, the
probability distribution $p({\mathbf{s}}|T)$ is given by
Eq.~(\ref{e:prior}). The function $p(\theta,\phi)$ is the expected
distribution of source directions.  For sources
that are mostly further than $\sim 30 \, \text{Mpc}$, the distribution
should be uniform on the sphere.

The integral in the expression (\ref{e:multiLikelihood3}) for the
likelihood function includes a sum over all
time-frequency windows ${\cal T}$ and all source directions ${\bf m}$.
However, it is nearly equivalent, and much easier, to adopt the
\emph{maximum} term in the sum as an approximation to the likelihood
function, since the largest term will dominate the sum when a signal
is present.  It is therefore sufficient to consider only 
a single time-frequency region ${\cal T}$ and fixed direction
$\mathbf{m}$ in the remainder of this section, with the understanding
that the detection statistic will include a maximization over these
variables \cite{note}.

Using arguments similar to those in Sec.~\ref{ss:derivation}, we can
perform the 
integral over signals $\mathbf{s}$.  In particular, the prior probability
in Eq.~(\ref{e:prior}) restricts $\mathbf{s}$ to lie in a vector space
${\mathcal{V}}_{\cal T}$ which contains only signals in the
time-frequency window ${\cal T}$.  This has the effect of replacing
$\mathbf{s}$ by ${\mathbf{s}}_\parallel$ in the inner products.
Moreover, the inner product 
$\lbrak \cdot ,\cdot \rbrak$ induces a natural inner product on the subspace
${\mathcal{V}}_{\cal T}$.  
The integral over angles can
be performed as in Eq.~(\ref{e:likelihood;A}) to show that
\begin{equation}
\Lambda(\vec{\mathbf{h}}) = \int \Gamma(V)e^{-A^2/2}
  (A{\mathcal{E}}^{1/2}/2)^{1-V} I_{V-1}(A{\mathcal{E}}^{1/2})
p(A) dA \label{e:multiLikelihood;A}
\end{equation}
where 
\begin{equation}
{\mathcal{E}} = \lbrak \tilde{h}^{\text{(eff)}}_{\parallel},
\tilde{h}^{\text{(eff)}}_{\parallel} \rbrak = 4{\mathrm{Re}}\sum_{k_1
\le k < k_2}  \tilde{h}^{\text{(eff)}}_k
\tilde{h}^{\text{(eff)}\ast}_k/S^{\text{(eff)}}_k .
\end{equation}
Here $\tilde{h}^{\text{(eff)}}_{\parallel}$ is the projection of the effective
strain into the time-frequency subspace ${\mathcal{V}}_{\cal T}$; the
second equality holds for a particular time-frequency window in which
the signal has duration 
$N\Delta t$ and is localized to the frequency band $k_1 \le k < k_2$.  The
amplitude $A$ is defined by the $A^2 = \lbrak \tilde{s}_{\parallel},
\tilde{s}_{\parallel} \rbrak$.  Since the right hand side of
Eq.~(\ref{e:multiLikelihood;A}) is a monotonically increasing function of
$\mathcal{E}$, the Neyman-Pearson theorem tells us that $\mathcal{E}$ provides
the optimal multi-instrument detection statistic.  Note that, as
mentioned above, this detection statistic includes an implicit
maximization over all source directions ${\bf m}$ and time-frequency
windows ${\cal T}$.  

\subsection{General networks of detectors}
\label{ss:multiGeneral}

When the network contains at least one instrument with a different orientation
to the others, it is necessary to discuss the two degrees of freedom, or
polarizations, of the gravitational wave signal.  We denote these two
independent signals as $s^+(t)$ and $s^\times(t)$, where the
definition is with respect to a radiation coordinate system associated
with the gravitational waves.  (See Appendix \ref{a:response} for a
detailed discussion of these and the other coordinate systems relevant
to this section.)  For the $A$th detector in the 
network, the gravitational wave strain is
\begin{equation}
s^A(t) = F_+^A s^+(t + \tau_A) + F_\times^A s^\times(t + \tau_A)
\label{e:plusCrossSignal}
\end{equation}
where $F^A_+$, $F^A_\times$ are the detector beam pattern functions and
$\tau_A = {\mathbf{m}} \cdot {\mathbf{x}}^A/ c$ is the time delay between the
origin of earth-fixed coordinates and the detector $A$ (located at
${\mathbf{x}}^A$) for a wave propagating in the direction ${\mathbf{m}}$.

The concept of effective strain
is particularly useful in the formal 
development of the multi-instrument detection statistic for
gravitational waves.  To introduce this
concept, consider the inner product 
\begin{equation}
(\vec{\mathbf{s}},\vec{\mathbf{s}}) = \sum_{A,B=1}^{n_d} 4 {\mathrm{Re}}
\sum_{k=0}^{\lfloor N/2\rfloor} \tilde{s}^A_{k}
({\mathbf{S}}^{-1}_k)^{AB} \tilde{s}^{B\, *}_{k}  .
\end{equation}
The DFT of the signal at the $A$th detector is
\begin{equation}
\tilde{s}_k^A = e^{2 \pi i \Delta_A({\mathbf{m}}) k /N}
(F^A_+ \tilde{s}_k^+ + F^A_\times \tilde{s}_k^\times)
\end{equation}
where $\Delta_A({\mathbf{m}}) = \tau_A/\Delta t$ is the discrete time delay,
$\Delta t$ is the sampling rate, and $\tilde{s}^{+,\times}$ are the DFTs of
the plus and cross polarizations of the signal defined in
Eq.~(\ref{e:plusCrossSignal}).  The inner product can be rewritten as
\begin{equation}
(\vec{\mathbf{s}},\vec{\mathbf{s}}) = \sum_{\alpha,\beta=+,\times} 4
{\mathrm{Re}} \sum_{k=0}^{\lfloor N/2\rfloor}
\tilde{s}^\alpha_{k} \Theta_{\alpha\beta}^k \tilde{s}^{\beta *}_{k}
\label{e:misMultiInnerProd}
\end{equation}
where 
\begin{equation}
\Theta_{\alpha\beta}^k = \sum_{A,B=1}^{n_d} e^{2 \pi i (\Delta_A -
\Delta_B) k /N} F^A_\alpha ({\mathbf{S}}^{-1}_k)^{AB} F^B_\beta  .
\end{equation}
We can now introduce a pair of effective strains, corresponding to the $+$ and
$\times$ gravitational wave signals for the network,  by
\cite{flanagan.e:1998a} 
\begin{equation}
\tilde{h}_k^\alpha =
\sum_{\beta=+,\times} \Theta^{\alpha\beta}_k \sum_{A,B=1}^{n_d}
F^A_\alpha e^{-2 \pi i \Delta_A k /N}  ({\mathbf{S}}^{-1}_k)^{AB}
\tilde{h}^{B}_k 
\end{equation}
where $\sum_{\beta=+,\times} \Theta^{\alpha\beta}_k \Theta_{\beta\gamma}^{k}
= \delta^\alpha_\gamma$.  In terms of the effective strains
$\tilde{h}_k^\alpha$ and the signals $s_{+,\times}$,   the likelihood ratio is
\begin{equation}
\Lambda(\vec{\mathbf{h}};\vec{\mathbf{s}}) 
= \exp[ \lbrak \tilde{h}, \tilde{s} \rbrak - \case{1}{2}
\lbrak \tilde{s},\tilde{s} \rbrak ]
\label{e:multiLikelihood4}
\end{equation}
where the inner product is now defined as
\begin{equation}
\lbrak \tilde{p}, \tilde{q} \rbrak = \sum_{\alpha,\beta=+,\times} 4
{\mathrm{Re}} \sum_{k=0}^{\lfloor N/2\rfloor}
\tilde{p}^\alpha_{k} \Theta_{\alpha\beta}^k \tilde{q}^{\beta *}_{k}  .
\label{e:innerproduct13}
\end{equation}
The effective strains and the inner product (\ref{e:innerproduct13})
depend on the 
direction $\mathbf{m}$ to the putative source.  Consequently the probability
that a signal with plus polarization $s^+$ and cross polarization $s^\times$
is present in the data stream is given by Eqs.~(\ref{e:multiLikelihood2}) and
(\ref{e:multiLikelihood3}) with $\Lambda(\vec{\mathbf{h}};\vec{\mathbf{s}})$
defined by Eq.~(\ref{e:multiLikelihood4}) where the measure
$p({\mathbf{s}}|{\cal T})$ on the space of signals is defined as follows.

The signal $\{s^+,s^\times\}$ now belongs to a $4V$-dimensional vector
space which is the tensor 
product of two copies of ${\mathcal{V}}_{\cal T}$.  Within this vector
space, all directions can also be considered to be equally likely.  These
assumptions about the signal $\{s_+, s_\times\}$ reduce to the assumption in
Sec.~\ref{ss:multiAligned} when the detectors are aligned.  Moreover, the
reasoning from that section can be readily applied here provided one
understands that the vector space of signals is now $4V$-dimensional and that
all angles and lengths are measured using the inner product 
(\ref{e:misMultiInnerProd}).  Thus, the integrals can be carried out in
much the same way to arrive at the excess power statistic for a
multiple-instrument network:
\begin{equation}
{\mathcal{E}} = \lbrak \tilde{h}_{\parallel},
\tilde{h}_{\parallel} \rbrak = \sum_{\alpha,\beta=+,\times}
4{\mathrm{Re}} \sum_{k_1 \le k < k_2}
\tilde{h}^{\alpha}_k \Theta_{\alpha\beta}^k \tilde{h}^{\beta\ast}_{k}.
\label{e:calEdef10}
\end{equation}
As before there is an implicit maximization
over time-frequency windows and source directions.

Since the effective strains are linear combinations of the outputs of each
of the detectors in the network, the statistic (\ref{e:calEdef10})
is a bilinear function of the outputs of all the detectors, containing
both auto-correlation terms from each detector individually and
cross-correlation terms between each pair of detectors.
It is the optimal statistic in Gaussian noise.  When the noise in the
instruments is non-Gaussian, it remains to be seen what is the best
strategy.  One obvious strategy is to simply omit the auto-correlation
terms in Eq.\ (\ref{e:calEdef10}) and retain only the cross-correlation
terms; the resulting statistic will share many of the nice features of
${\cal E}$ and be more robust against non-Gaussian noise bursts.
The real challenge is to derive the optimal statistic in the presence of
uncorrelated noise bursts which are Poisson distributed in time.  It is likely
that the model introduced in Ref.~\cite{creighton.j:1999b} can be used to
address this issue.

\acknowledgements
We are grateful to Bruce Allen, Sam Finn, Soumya Mohanty, Julien Sylvestre and
Kip Thorne for helpful discussions.
This work was supported by National Science Foundation awards
PHY~9970821,     
PHY~9722189,     
PHY~9728704,     
PHY-9407194,     
and Phy-9900776. 
EF acknowledges the support of the Alfred P. Sloan foundation.

\appendix

\section{Related detection statistics}
\label{s:related}

In this appendix we discuss some detection statistics that can be
obtained from the Bayesian formalism discussed in Sec.\ \ref{s:bayes_analysis}
starting from different prior PDFs for signals ${\bf s}$.

\subsection{Known signal spectrum}
\label{ss:knownSpec}

Suppose that one knows, in addition to the duration and frequency band, the
spectrum of the expected signal, but that one does not know the phase
evolution. 
Let us adopt the Fourier basis (assuming that the autocorrelation matrix is
reasonably close to diagonal in this basis) and assume that the noise is
stationary and Gaussian.  Then the likelihood ratio is
\begin{equation}
\label{e:likelihoodKnownSpec;s}
  \Lambda({\mathbf{h}};{\mathbf{s}})
  = \exp\biggl[4{\textstyler{\sum_{k=0}^{\lfloor N/2\rfloor}}}
  (|\tilde{h}_k||\tilde{s}_k|\cos\phi_k - \case{1}{2}|\tilde{s}_k|^2)/S_k
  \biggr]
\end{equation}
where $\phi_k$ represents the relative phase difference between the data and
the expected signal in the $k$th frequency bin.  Since the signal phases are
considered unknown, we should integrate out these angles to obtain the
integrated likelihood ratio
\begin{equation}
\label{e:likelihoodKnownSpec;P}
  \Lambda({\mathbf{h}};\{P_k\},A)
  = \prod_{k=0}^{\lfloor N/2\rfloor} 2\pi e^{-A^2P_k/S_k}
  I_0(2^{3/2}AP_k^{1/2}|\tilde{h}_k|/S_k)
\end{equation}
where the one-sided signal spectrum is given by
$\case{1}{2}A^2P_k=|\tilde{s}_k|^2$ with
$2\sum_{k=0}^{\lfloor N/2\rfloor}P_k/S_k=1$.

In the limit of weak signal amplitude $A$, we can approximate the likelihood
ratio of Eq.~(\ref{e:likelihoodKnownSpec;P}) by its expansion in powers of
$A$.  The first non-trivial term is the \emph{locally
optimal}~\cite{Kassam.S:1988} detection statistic
\begin{equation}
\label{e:locallyOptimalKnownSpec}
  \left.\frac{d^2\ln\Lambda({\mathbf{h}};\{P_k\},A)}{dA^2}\right|_{A=0}
  = \mbox{(const)} + 4\sum_{k=0}^{\lfloor N/2\rfloor}P_k|\tilde{h}_k|^2/S_k^2.
\end{equation}
This statistic is the weighted average of the detector output power in each
frequency bin.  Unfortunately, it is not possible to get simple expressions
for the false alarm and false dismissal probabilities for this statistic; one
needs to use numerical methods to obtain these given a known signal power
spectrum $\{P_k\}$.

\subsection{Non-Gaussian noise}
\label{ss:nonGaussian}

It is unlikely that a gravitational wave detector will produce purely
stationary and Gaussian noise.  In the case that the detector noise
distribution is known, we can obtain a detection statistic for unknown
signals using the Bayesian methodology.  Unfortunately the most
general noise distribution contains many free functions and 
will not be known in practice.  However, constructing
simple analytic non-Gaussian noise models and the associated detection
statistics can give us insight into what kind of statistics to
try out with real detectors.

One such simple model is as follows.  We assume that the detector noise is
stationary, and, as before, that each frequency bin in the Fourier
basis is uncorrelated.  Let make the 
additional assumption that the power in each frequency bin is
independentally distributed, while the phases of each frequency bin are
uniform and independent.  Then
\begin{equation}
  p({\mathbf{n}}) d^N{\mathbf{n}} =
  {\textstyler{\prod_{k=1}^{\lfloor(N-1)/2\rfloor}}}
  f_k(|\tilde{n}_k|^2/S_k) \frac{d|\tilde{n}_k|^2}{S_k}
  \frac{d\arg\tilde{n}_k}{2\pi}
\end{equation}
where $f_k(x)$ are known non-exponential probability distribution functions
(exponential functions would correspond to Gaussian noise).  Here we have
omitted the DC (and a possible Nyquist) component.  The likelihood ratio is
\begin{eqnarray}
  \Lambda({\mathbf{h}};{\mathbf{s}})
  &=& \prod_{k=1}^{\lfloor(N-1)/2\rfloor}
  \frac{f_k(|\tilde{h}_k-\tilde{s}_k|^2/S_k)}{f_k(|\tilde{h}_k|^2/S_k)}
  \nonumber\\
  &=& \prod_{k=1}^{\lfloor(N-1)/2\rfloor} \biggl\{
  1-2\frac{{\mathrm{Re}}(\tilde{h}_k\tilde{s}_k^\ast)}{S_k}
  \frac{f_k'(|\tilde{h}_k|^2/S_k)}{f_k(|\tilde{h}_k|^2/S_k)}
  \nonumber\\ &&
  + \biggl[ 2\biggl(\frac{{\mathrm{Re}}(\tilde{h}_k\tilde{s}_k^\ast)}{S_k}
  \biggr)^2
  \frac{f_k''(|\tilde{h}_k^2/S_k)}{f_k(|\tilde{h}_k^2/S_k)}
  \nonumber\\ && \quad
  + \frac{|\tilde{s}_k|^2}{S_k}
  \frac{f_k'(|\tilde{h}_k|^2/S_k)}{f_k(|\tilde{h}_k^2/S_k)} \biggr]
  + O(|\tilde{s}_k|^3) \biggr\}.
\end{eqnarray}
We have expanded the likelihood ratio in powers of the (presumed small) signal
in order to construct the locally optimal detection
statistic~\cite{Kassam.S:1988}.

To compute the integrated likelihood function we need to integrate over our
prior knowledge of signals.  Let us suppose that we do not know the signal
phase evolution; then we can integrate over the unknown phases
$\arg\tilde{s}_k$ in each frequency bin.  We find
\begin{equation}
\label{e:locallyOptimalNonGaussian}
  \Lambda({\mathbf{h}};\{P_k\},A) =
  1 + \frac{A^2}{2}\sum_{k=1}^{\lfloor(N-1)/2\rfloor}\frac{|\tilde{s}_k|^2}{S_k}
  g_k(|\tilde{h}_k|^2/S_k) + O(A^4)
\end{equation}
where $\case{1}{2}A^2P_k=|\tilde{s}_k|^2$ and
\begin{equation}
  g_k(x) = [xf''_k(x)+f'_k(x)]/f_k(x).
\end{equation}
For Gaussian noise, $f_k(x)=e^{-x}$ and $g_k(x)=x-1$ for all $k$, which gives
essentially the same detection statistic as in
Eq.~(\ref{e:locallyOptimalKnownSpec}).  For a
probability distribution with tails that decrease more slowly in the $k$th
bin, e.g., $f_k(x)\propto(1-x/2)^{-2}$, then we have $g_k(x)=(x-1)/(1-x/2)^2$,
which increases with $x$ up to $x=2$, and then decreases for larger values of
$x$.  Thus, large amounts of excess power in the $k$th bin are
\emph{suppressed}.

When the signal is known to be band-limited to frequency bins
$k_1 \le k < k_2$, but we have no reason to believe that any
particular bin in 
the band will contribute more to the overall signal-to-noise ratio than any
other bin, then we obtain the locally optimal statistic by assuming a uniform
weighting of the terms in Eq.~(\ref{e:locallyOptimalNonGaussian}).  Thus the
locally optimal statistic is
\begin{equation}
  {\textstyler{\sum_{k_1 \le k < k_2}}} g_k(|\tilde{h}_k|^2/S_k).
\end{equation}
In the case of Gaussian noise this is the excess power statistic; for noise
models with larger tails, the components of the sum are attenuated if they have
large power.

\section{Multidetector amplitudes}
\label{a:response}

In Sec.~\ref{ss:multiGeneral}, we discussed the detection of burst signals
using multiple detectors.  When the detectors are not aligned, one needs the
response functions of each detector in the network to a gravitational wave
signal from a given sky position.  Here we define reference coordinates to
which we refer each detectors response.  Consider a coordinate system fixed at
the center of the earth.  In terms of latitude and longitude
$\{\varphi,\lambda\}$, the coordinate axes are oriented so that the $x$-axis
pierces the earth at $\{000,000\}$, the $y$-axis pierces the earth at
$\{000,090^\circ\text{E}\}$, and the $z$-axis pierces the earth at
$\{090^\circ\text{N},000\}$.  We denote the location of a source on the
celestial sphere by standard spherical polar coordinates $\{\theta, \phi\}$
measured with respect to this earth fixed frame.  A fiducial signal
$\mathbf{s}$ comes from a sky position with right ascension
$\alpha=\phi+\mbox{GMST}$ (GMST is the Greenwich mean sidereal time of arrival
of the signal), declination $\delta=\pi/2-\theta$, and has polarization angle
$\psi$---the angle (counter-clockwise about the direction of propagation) from
the line of nodes to the $\mathbf{X}$-axis of the signal coordinates.  In
particular, this gravitational wave signal can be represented by a tensor
\begin{equation}
s_{ij} = s^+ ({\mathbf{e}}_+)_{ij} + s^\times ({\mathbf{e}}_\times)_{ij}
\end{equation}
where the polarization tensors are given by
\begin{eqnarray}
({\mathbf{e}}_+)_{ij} &=& ( {\mathbf{X}} \otimes {\mathbf{X}} -
{\mathbf{Y}} \otimes {\mathbf{Y}})_{ij} \\ ({\mathbf{e}}_\times)_{ij} &=&
( {\mathbf{X}} \otimes {\mathbf{Y}} + {\mathbf{Y}} \otimes
{\mathbf{X}})_{ij}  .
\end{eqnarray}
The vectors $\mathbf{X}$ and $\mathbf{Y}$ are the axes of the wave-frame,
given explicitly by
\begin{eqnarray}
{\mathbf{X}} &=& (\sin \phi \, \cos \psi - \sin \psi \, \cos \phi \, \cos
\theta)\; {\mathbf{i}} \nonumber \\
&& \mbox{\vspace{0.5in}}
-(\cos \phi \, \cos \psi  + \sin \psi \, \sin \phi \, \cos \theta)\;
{\mathbf{j}} +
\sin \psi \, \sin \theta \; {\mathbf{k}} \label{e:waveX}\\
{\mathbf{Y}} &=& (- \sin \phi \, \sin \psi - \cos \psi \, \cos \phi \, \cos
\theta)\;  {\mathbf{i}} \nonumber \\
&& \mbox{\vspace{0.5in}}
  +( \cos \phi \, \sin \psi - \cos \psi \, \sin \phi \, \cos \theta) \;
  {\mathbf{j}}
 + \sin \theta \, \cos \psi\;  {\mathbf{k}} \;  \label{e:waveY}
\end{eqnarray}
where the polarization angle $\psi$ is defined above, and $\mathbf{i}$,
$\mathbf{j}$ and $\mathbf{k}$ are unit vectors along the $x$, $y$ and $z$-axes
respectively.  Note, we use a right handed coordinate system in which the
vector ${\mathbf{Z}}= {\mathbf{X}} \wedge {\mathbf{Y}}$ points in the
direction from the source towards the detector.  The waveforms in Refs.
\cite{will.c:1996,blanchet.l:1996} are referred to these coordinates; Thorne
uses a different definition in Ref.~\cite{300yrs}.

One can characterize the response of an interferometer on the surface of the
earth to the impinging gravitational wave using another tensor $\mathbf{D}$
given by
\begin{equation}
{\mathbf{D}}_{ij} = \case{1}{2} ({\mathbf{n}}^x \otimes {\mathbf{n}}^x -
{\mathbf{n}}^y \otimes {\mathbf{n}}^y)_{ij} \label{e:detectorResponse}
\end{equation}
where ${\mathbf{n}}^x$ and ${\mathbf{n}}^y$ are unit vectors along the $x$ and
$y$ arms of the interferometer respectively.  For a given interferometer $A$,
it is now straightforward to compute the response
\begin{equation}
s^A = \sum_{i,j=1}^3 D^A_{ij} s_{ij} \label{e:generalSignal}
\end{equation}
and to extract the response functions $F^A_{+,\times}$ by comparing the result
with the formula:
\begin{equation}
s^A= F^A_+ s^+ + F^A_\times s^\times  .
\end{equation}
For a detector having its arms aligned with the coordinate axes at the center
of the earth,  we find
\begin{eqnarray}
F_+ &=& - \case{1}{2} ( 1 + \cos^2 \theta ) \, \cos 2 \phi \, \cos 2\psi \, 
- \cos \theta \, \sin 2 \phi \, \sin 2\psi \\
F_\times &=& + \case{1}{2}  ( 1+ \cos^2 \theta)\, \cos 2\phi \, \sin 2 \psi -
\cos \theta \, \sin 2\phi \, \cos 2 \psi 
\end{eqnarray}

Finally, the response $\mathbf{D}$ of the various detectors around the world
can be determined using the latitude North $\varphi$, longitude East
$\lambda$, and arm orientations $(\psi_{x,y},\omega_{x,y})$ where $\psi_{x,y}$
are the azimuths (North of East) of the $x$ and $y$ arms and $\omega_{x,y}$
are the tilts of the $x$ and $y$ arms above the horizontal defined by the
WGS-84 earth model~\cite{LIGO.T980044-08}.  This model is an oblate ellipsoid
with semi-major axis $a=6\,378\,137\,\text{m}$ and semi-minor axis
$b=6\,356\,752.314\,\text{m}$.  The position
${\mathbf{x}}=x\;{\mathbf{i}}+y\;{\mathbf{j}}+z\;{\mathbf{k}}$ of a detector
at a given latitude $\varphi$, longitude $\lambda$, and elevation $h$ above
(normal to) the surface is given by
\begin{eqnarray}
  x &=& [R(\varphi) + h]\cos\varphi\cos\lambda \\
  y &=& [R(\varphi) + h]\cos\varphi\sin\lambda \\
  z &=& [(b^2/a^2)R(\varphi) + h]\sin\varphi
\end{eqnarray}
where $R(\varphi)=a^2(a^2\cos\varphi+b^2\sin\varphi)^{-1/2}$ is the local
radius of the earth.  At this position, the unit vectors pointing East, North,
and Up are
\begin{eqnarray}
  {\mathbf{e}}_\lambda &=& -\sin\lambda\;{\mathbf{i}} +
    \cos\lambda\;{\mathbf{j}} \\
  {\mathbf{e}}_\varphi &=& -\sin\varphi\cos\lambda\;{\mathbf{i}} -
    \sin\varphi\sin\lambda\;{\mathbf{j}} + \cos\varphi\;{\mathbf{k}} \\
  {\mathbf{e}}_h &=& \cos\varphi\cos\lambda\;{\mathbf{i}} +
    \cos\varphi\sin\lambda\;{\mathbf{j}} + \sin\varphi\;{\mathbf{k}}
\end{eqnarray}
respectively.  The unit vector along the $x$ arm is then given by
\begin{equation}
  {\mathbf{n}}^x = \cos\omega_x\cos\psi_x\;{\mathbf{e}}_\lambda +
  \cos\omega_x\sin\psi_x\;{\mathbf{e}}_\varphi + \sin\omega_x\;{\mathbf{e}}_h
\end{equation}
and similarly for the $y$ arm.  For completeness, we list these vectors
${\mathbf{n}}^x$ and ${\mathbf{n}}^y$ for each of the interferometers in
Table~\ref{t:detectors}.  For the two LIGO interferometers, these vectors are
provided in~\cite{LIGO.T980044-08}.  For the other interferometers we used the
values in note~\cite{sites} (with tilt angles $\omega=0$), or the values given
in Ref.~\cite{allen.b:1996a} (with elevations $h=0$ and tilt angles
$\omega=0$).


\clearpage
\widetext
\begin{table}
\label{t:detectors}
\caption{The locations $\mathbf{x}$ (the coordinates are in
meters) and direction vectors $\{{\mathbf{n}}^x , {\mathbf{n}}^y\}$ for the
various interferometers around the world based on the the data
in~\protect\cite{allen.b:1996a,sites} and the ellipsoidal model described
in~\protect\cite{LIGO.T980044-08}.  Note that the arm orientations reported
by Allen~\protect\cite{allen.b:1996a} for the two LIGO interferometers do
not correctly represent the angles between the northing and the arms at the
two sites; we have therefore stated the official LIGO arm orientation
vectors.  For VIRGO, GEO-600, and TAMA-300, see~\protect\cite{sites}.
For the first five interferometers, the results are based entirely on the
numbers reported in Allen.}
\begin{tabular}{llccc}
Project & Location & ${\mathbf{n}}^x$ &
${\mathbf{n}}^y$ & ${\mathbf{x}}$ (
$\times 10^6$ m)\\
\hline
CIT     & Pasadena, CA, USA &
          \{$-$0.2648,$-$0.4953,$-$0.8274\} &
          \{$+$0.8819,$-$0.4715,$+$0.0000\} &
          \{$-$2.490650,$-$4.658700,$+$3.562064\} \\
MPQ     & Garching, Germany &
          \{$-$0.7304,$+$0.3749,$+$0.5709\} &
          \{$+$0.2027,$+$0.9172,$-$0.3430\} &
          \{$+$4.167725,$+$0.861577,$+$4.734691\} \\
ISAS-100& Tokyo, Japan &
          \{$+$0.7634,$+$0.2277,$+$0.6045\} & 
	  \{$+$0.1469,$+$0.8047,$-$0.5752\} &
 	  \{$-$3.947704,$+$3.375234,$+$3.689488\} \\
TAMA-20 & Tokyo, Japan &
          \{$+$0.7727,$+$0.2704,$+$0.5744\}&
 	  \{$-$0.1451,$-$0.8056,$+$0.5744\}&
	  \{$-$3.946416,$+$3.365795,$+$3.699409\} \\
Glasgow & Glasgow, UK &
          \{$-$0.4534,$-$0.8515,$+$0.2634\} &
	  \{$+$0.6938,$-$0.5227,$-$0.4954\} &
	  \{$+$3.576830,$-$0.267688,$+$5.256335\} \\
\hline
TAMA-300& Tokyo, Japan &
          \{$+$0.6490,$+$0.7608,$+$0.0000\} &
	  \{$-$0.4437,$+$0.3785,$-$0.8123\} &
          \{$-$3.946409,$+$3.366259,$+$3.699151\} \\
GEO-600 & Hannover, Germany &
          \{$-$0.6261,$-$0.5522,$+$0.5506\} & 
	  \{$-$0.4453,$+$0.8665,$+$0.2255\} &  
	  \{$+$3.856310,$+$0.666599,$+$5.019641\} \\
VIRGO   & Pisa, Italy &
          \{$-$0.7005,$+$0.2085,$+$0.6826\} &  
	  \{$-$0.0538,$-$0.9691,$+$0.2408\} &
          \{$+$4.546374,$+$0.842990,$+$4.378577\} \\
LIGO    & Hanford, WA, USA &
          \{$-$0.2239,$+$0.7998,$+$0.5569\} & 
	  \{$-$0.9140,$+$0.0261,$-$0.4049\} &      
	  \{$-$2.161415,$-$3.834695,$+$4.600350\} \\
LIGO    & Livingston, LA, USA &
          \{$-$0.9546,$-$0.1416,$-$0.2622\} & 
	  \{$+$0.2977,$-$0.4879,$-$0.8205\} & 
          \{$-$0.074276,$-$5.496284,$+$3.224257\} \\
\end{tabular}
\end{table}

\clearpage
\narrowtext

\newpage

\newpage

\newpage 

\newpage

\newpage

\end{document}